\documentclass[12pt]{iopart}
\usepackage{graphicx}
\expandafter\let\csname equation*\endcsname\relax
\expandafter\let\csname endequation*\endcsname\relax
\usepackage{amsmath}
\usepackage{mathabx}
\usepackage{tikz}
\usetikzlibrary{shapes,arrows,positioning}
\tikzset{near start abs/.style={xshift=1cm}}
\usepackage{lscape}
\usepackage{varwidth}
\usepackage{color,soul}
\newcommand{\tilderho}{\tilde{\rho}({\mathbf x})}
\newcommand{\tildeh}{\tilde{h}({\mathbf x})}
\newcommand{\E}{{\rm E}}
\newcommand{\Var}{{\rm Var}}
\newcommand{\Cov}{{\rm Cov}}

\begin{document}

\title[Geodesy and modified gravity]{Interpretation of geodesy experiments in non-Newtonian theories of gravity}

\author{Joel Berg\'e$^1$, Philippe Brax$^2$, Martin Pernot-Borr\`as$^1$$^3$, Jean-Philippe Uzan$^4$$^5$}

\address{$^1$ DPHY, ONERA, Universit\'e Paris Saclay, F-92322 Ch\^atillon, France}
\address{$^2$ Institut de Physique Th\'eorique, Universit\'e Paris-Saclay, CEA, CNRS, F-91191 Gif-sur-Yvette Cedex, France}
\address{$^3$ Sorbonne Universit\'e, CNRS, Institut d'Astrophysique de Paris, IAP, F-75014 Paris, France}
\address{$^4$ CNRS, Institut d'Astrophysique de Paris, IAP, F-75014 Paris, France}
\address{$^5$ Institut Lagrange de Paris, 98 bis, Bd Arago, 75014 Paris, France}
\ead{joel.berge@onera.fr}
\vspace{10pt}
\begin{indented}
\item[]July 2018
\end{indented}

\begin{abstract}
The tests of the deviations from Newton's or Einstein's gravity in the Earth neighbourhood are tied to our knowledge of the shape and mass distribution of our planet. On the one hand estimators of these ``modified" theories of gravity may be explicitly Earth-model-dependent whilst  on the other hand the Earth gravitational field would act as a systematic error. We revisit  deviations from Newtonian gravity described by a Yukawa interaction that can arise from the existence of a finite range fifth force. We show that the standard multipolar expansion of the Earth gravitational potential can be generalised. In particular, the multipolar coefficients  depend on the distance to the centre of the Earth and are therefore not universal to the Earth system anymore. This  offers new ways of constraining such  Yukawa interactions and demonstrates explicitly the limits of the Newton-based interpretation of geodesy experiments. In turn, limitations from geodesy data restrict the possibility of testing gravity in space.  The gravitational acceleration is  described in terms of spin-weighted spherical harmonics allowing  us to obtain the perturbing force entering the Lagrange-Gauss secular equations. This  is then used to discuss the correlation between geodesy and modified gravity experiments and the possibility to break their degeneracy. Finally we show that, given the existing constraints, a Yukawa fifth force is expected to be sub-dominant in satellite dynamics and space geodesy experiments, as long as they are performed at altitudes greater than a few hundred kilometres. Gravity surveys will have to gain at least two orders of magnitude in instrumental precision before satellite geodesy could be used to improve the current constraints on modified gravity.
\end{abstract}

%
\noindent{\it Keywords}: Experimental Gravitation, Modified Gravity
%

\submitto{\CQG}
%
%
%

\section{Introduction}

The efforts to test Newton and Einstein gravity have been continuous in the last hundred years and lie at the crossroads between theoretical and experimental physics, laboratory and space physics.  Celestial mechanics has historically been  crucial  in that respect, motivated mostly by the imperfect  understanding of the shape of the Earth,  the stability of the Solar system and the long lasting Newtonian problem of the anomalous drift of the perihelion of Mercury. A main difficulty arises from the fact that gravity is a long range interaction that cannot be screened. Hence, the knowledge of our environment (Earth gravitational field and its evolution, Solar system structure, cosmological model) is a limitation to these tests. In that respect, the developments of dedicated satellite missions have brought new insights on both possible deviations from General Relativity (GR)  and the Earth gravitational field.

Today, GR is well-tested on local scales~\cite{Will:1993ns,will14} whilst the need to improve the existing constraints is partly motivated by cosmology. The accelerated cosmic expansion and other evidences, such as the dynamics of spiral galaxies, have led to the conclusion that there should exist a dark sector, composed of dark matter and dark energy, representing respectively 26\% and 70\% of the energy budget of the universe. This dark sector can also be interpreted as a sign that GR may not be a good description of gravity on cosmological scales and on low acceleration regimes. Many extensions of GR have been proposed (see e.g. Refs.~\cite{adelberger03,clifton12, joyce15} for reviews) and many tests of GR and of Einstein's equivalence principle on cosmological scales have been designed to test them~\cite{Uzan:2000mz,Uzan:2003zq,Uzan:2006mf,Uzan:2010ri,Jain:2013wgs}. No deviations from GR have been detected so far (see e.g. Refs.~\cite{will14,adelberger03, jain10,safronova17,delva17} for general reviews of laboratory and cosmological scales tests).

Concerning the recent experimental tests of GR, let us mention the Lense-Thirring effect \cite{everitt11, ciufolini13a, ciufolini13b}, the pericentre anomaly \cite{iorio02a, lucchesi03, lucchesi10, lucchesi14, li14}, the gravitational redshift \cite{bertotti03}, the universality of free fall  \cite{williams04,schlamminger08,wagner12,touboul17,viswanathan18} and the constancy of fundamental constants~\cite{Uzan:2002vq,Uzan:2010pm,Uzan:2004qr}, the last two involving  Einstein's equivalence principle. To these standard tests, let us add the new window opened by the detection of gravitational waves~\cite{abbott16}. In particular, the detection of an electromagnetic counterpart to the gravitational-wave signal emitted by a binary neutron star merger~\cite{abbott17} put severe constraints on a whole class of alternatives to GR~\cite{baker17,creminelli17,ezquiaga17,sakstein17}.

Among all the extensions of GR, scalar-tensor theories~\cite{damour92}, in which a scalar long range interaction that may be composition dependent, is added to the standard spin-2 interaction mediated by the gravitons, are still among the open alternatives. In particular, if this scalar is light, they may enjoy sizeable cosmological imprints. As a long range fifth force  would then appear on Solar system scales, they need to include a screening or a decoupling mechanism~\cite{vainshtein72,Damour:1992kf,damour94,khoury04a, khoury04b, babichev09,hinterbichler10, brax13}. While the parameter space of these models has been severely reduced (see e.g. Ref.~\cite{burrage18} for up-to-date tests), they remain ideal candidates for extensions of GR. Even if the scalar field is heavy on Solar system scales, it is still responsible for a fifth force described, in the Newtonian regime, by a Yukawa potential (see e.g. the Supplemental material of Ref.~\cite{berge18} and references therein). Many constraints on the mass and  the amplitude of this extra-potential have been obtained so far (e.g. Refs.~\cite{adelberger03,fischbach99} and references therein, and Refs.~\cite{kapner07,masuda09,sushkov11,klimchitskaya14,berge18} for more recent works).

The goal of this article is to revisit the constraints on such a Yukawa interaction drawn from the analysis of geodetic data. As already emphasised, it is a tautology to say that local GR tests are limited by our knowledge of the Earth gravitational field. Nevertheless, there have been extensive studies under the assumption of Newton's gravity whilst the tests of Yukawa gravity have all been performed assuming at best a spherical and homogeneous Earth, but most often, assuming that the Yukawa interaction is sourced by a point-like Earth. We develop a method to describe the effects of such a modified gravity on the orbits of dedicated satellites in a realistic description of the Earth. Clearly, in that case our ignorance of the properties of the fifth force does limit our reconstruction of the property of the mass distribution of the Earth, while the latter limits the constraints on this fifth force. We propose to analyze these interactions and provide tools to test GR in our terrestrial neighbourhood.

The shape and mass distribution of the Earth, and their variability,  have so far been reconstructed from local measurements of the  gravitational field (on-ground or airborne) and  global satellite models of the full gravitational field. Recent satellite geodesy missions have allowed geophysicists to map the Earth gravity model with an exquisite precision: e.g. GOCE \cite{rummel11, pail11} or GRACE \cite{tapley04, tapley05, reigber05} and combinations of (satellite) missions~\cite{pail10, mayer06}. GOCE and GRACE provide measurements of the spherical harmonics coefficients up to degree and order 250, whereas the EGM2008 model goes up to degree and order 2159 \cite{pavlis12}.

The uncertainties on the shape of the Earth add up to other systematic errors (such as Solar radiation pressure, atmospheric drag, Earth tides, Earth magnetic field, thermal instabilities --for discussions of systematics in both laboratory and space, see e.g. Refs.~\cite{lucchesi14,touboul17,hoyle04}). Then, they must be either shielded or corrected for during the data analysis process (see e.g. Refs.~\cite{touboul17, hoyle04,fischbach86, toth18}). This article focuses on satellite tests of gravity so that the main sources of gravitational error come from the zonal terms, and especially the first one, $J_2$ (which describes the Earth flattening)~\cite{ciufolini13a, lucchesi14}.  Before the advent of the precise satellite measurements from GRACE and GOCE, the large uncertainty on $J_2$ was considered a show-stopper for precise tests of gravity. Techniques were then elaborated to cancel its effect. For instance, by empirically combining the perigee shift and precession of the line of nodes of LAGEOS and LAGEOS II, it was shown that the contribution of $J_2$ (and the associated error) to the perigee shift and to the Lense-Thirring effect could be cancelled \cite{ciufolini96}. The GRACE and GOCE missions changed the situation  thanks to their remarkably precise measurements, giving the parameter $J_2$ to  a $10^{-8}$ relative precision level  when combined with LAGEOS data. In the case of the perigee shift measurement of the LAGEOS II satellite, Lucchesi \& Peron \cite{lucchesi14} evaluate that using the errors on $J_2$ provided by the EIGEN-GRACE02S gravitational field model \cite{reigber05} allows for a percent level test of GR's perigee shift with no further empirical correction.

However, correcting for the shape of the Earth when testing gravity in space relies on two pillars: (i) a model of the Earth gravitational field and (ii) accurate and precise values of the coefficients of the model. To the best of our knowledge, the model is always described as a spherical harmonics expansion derived from Laplace equation to solve for the Newtonian gravitational field sourced by the shape of the Earth. The values of the spherical harmonics coefficients are provided by Earth gravity surveys, such as GRACE, GOCE, LAGEOS, or local on-ground surveys.

The evaluation of the accuracy of coefficients estimator and of robust uncertainties is a highly non-trivial part of the data analysis needed to make a model of the gravitational field.
Errors on spherical harmonic coefficients are commonly separated between formal and calibrated errors \cite{lucchesi14,reigber05}. Formal errors come from the data regression method and mainly include statistical errors as well as possible numerical uncertainties linked to the data analysis method itself. For instance, because of its Sun-synchronous orbit, GOCE never flew over the poles; the resulting polar gaps (whereby no data can constrain the spherical harmonics model in the polar regions) causes the least-square regression on spherical harmonics coefficients to be ill-conditioned, thus requiring a regularization technique. With no regularization, estimating the (near)-zonal terms is particularly difficult. These coefficients come with large error bars; after regularization, the error bars can be seen to shrink \cite{pail11, metzler05} (for $J_2$, the error shrinks from a few $10^{-9}$ to a few $10^{-12}$).  However, there does not seem to be  any investigation about the possible bias introduced by the regularization technique.

Under the Newtonian gravity hypothesis (i.e. the static part of spherical harmonics coefficients should be consistent between different data subsets along the experiment's time span, or between different experiments), formal errors are a posteriori calibrated to account for systematic errors: for a single satellite model, subset solutions are generated from data covering different time periods, and the scattering of subset solutions is used as the calibrated error (see e.g. Ref.~\cite{reigber05} for GRACE). The same method is applied to calibrate multi-satellite models, where an upper bounds for the systematic errors is derived from the difference between estimates of several satellite data \cite{lucchesi14}. In this case, it is implicitly assumed that any tension between different data sets comes from imperfectly controlled systematic errors. Although this is true if the underlying hypothesis (the Earth gravity is described by Newton's theory) is true, any tension may also provide a smoking gun for physics beyond Newton's inverse square law and GR. Indeed, a modified gravity model may very well predict non-universal spherical harmonics coefficients, e.g. coefficients whose value depends on the distance to the centre of the Earth (in this paper, we show that it is indeed the case). Along this line, it should be noted that despite very precise measurements of the static $J_2$ zonal term, the GRACE-only, GOCE-only and EIGEN-6C (combining LAGEOS, GOCE, GRACE and ground measurements) models provide inconsistent values (as was already noted by Wagner \& McAdoo \cite{wagnermcadoo12}), which differ by at least 700 $\sigma$; see Table~\ref{t.tab1}.

\begin{table}
\caption{Constraints on the $J_2$ parameter by several experiments.}
\begin{center}
\begin{tabular}{ccc}
\hline
GRACE & $J_2=1.0826354309122197\times10^{-3}\pm3.5263625612834223\times10^{-12}$ & \cite{mayer06}\\
GOCE & $J_2=1.0826265326404513\times10^{-3}\pm1.2127946116555258\times10^{-11}$ & \cite{pail11}\\
 EIGEN-6C & $J_2=1.0826263376893369\times10^{-3}\pm2.477786925867517\times10^{-13}$ &\cite{shako14}\\
\hline
\end{tabular}
\end{center}\label{t.tab1}
\label{default}
\end{table}%

Whether this tension is due to largely underestimated errors, to biases introduced by regularization techniques, to uncontrolled systematics, to inconsistent data sets, or to new physics beyond GR is not clear. However, it should invite us to  extreme caution when using gravity surveys and geodesy results to model and correct for the Earth gravitational field when testing GR or looking for deviations to Newton's inverse square law.

This article investigates the effects of modified gravity on the Earth gravitational field and our ability to reconstruct the shape of the Earth and, in turn, the effect of an imperfect knowledge of the Earth gravitational field on searches for modified gravity. As explained, we base our discussion on phenomenological deviations from Newton gravity described by a Yukawa potential.

In particular, we shall show that although we can still describe the Earth gravitational field with a spherical harmonics expansion, a Yukawa interaction modifies the meaning of the expansion coefficients. They mix  properties of the Earth and of gravity and get an explicit dependence on the distance to the centre of the Earth. As a consequence, they are not simply related to the Earth geometry any more, and should not be used to map the Earth mass distribution and geoid. For instance, the $J_2(r)$ zonal term does not only describe the Earth flattening, but is impacted by the Yukawa interaction. Furthermore, we should not expect coefficients measured by different satellites at different altitude (or even by a single satellite at different times, provided that satellite's orbit is not circular) to be consistent; combining different data sets should also be performed with great care.

Therefore, using geodesy results derived under the assumption that no deviation to GR (or to Newton's law) exists is prone to errors when constraining modified gravity, just because the Earth gravity model used to correct for the Newtonian contribution may be incorrect. This may be the case if using (possibly inconsistent) multi-satellite models, or a model set with a satellite at an altitude other than the altitude of the gravity test. The underlying question is that of the model to use. When looking for modified gravity in terms of a Yukawa interaction, two parameters are added to the Newtonian gravity sector (the strength and range of the interaction), {\it de facto} changing the model --which is not simply Newtonian any more. Using geodesy results derived assuming a simple Newtonian model must then be seen as inconsistent with the task at hand, and will introduce biases and uncertainties that must be quantified and accounted for in the modified gravity constraints.

The way out of this difficulty is, as usual, to set all analyses within the same theoretical framework to ensure consistency. The Earth gravitational field should be measured under the assumption that a Yukawa interaction may exist. The Earth gravitational field models would then explicitly contain information about the Yukawa interaction, either explicit or marginalised upon. In the former case, they would provide constraints on modified gravity; in the latter case, their estimated coefficients would have larger uncertainty, but would be unbiased and could safely be used by modified gravity experiments.

This paper is organised as follows. In Sect. \ref{sect_general} and \ref{sect_emodel}, we derive the spherical harmonic expansion of the Earth gravitational field in presence of a Yukawa interaction and give expressions for the gravitational acceleration and for the Gauss-Lagrange equations of motion. Sect. \ref{sect_impact} provides a general discussion of the entanglement between geodesy and modified gravity measurements, and order-of-magnitude estimates derived with a simple Earth model are given in Sect. \ref{sect_ofm}. This formalism provides a consistent framework to derive constraints on fifth force from space-borne experiments.

\section{Earth gravity in presence of a Yukawa potential} \label{sect_general}

\subsection{Gravitational potential}

Among the various ways to modify Newton's gravity, the introduction of a Yukawa potential describes the effect of an extra-massive degree of freedom that can appear, e.g. in scalar-tensor gravity~\cite{will14}. Assuming that the coupling of this new degree of freedom to the standard model fields is universal, the associated potential created by a point-mass source of mass $M$ at a distance $r$ is
\begin{equation} \label{eq_upointmass}
U_{\rm pm}(r) = -\frac{GM}{r} \left[1 + \alpha \exp\left(-\frac{r}{\lambda}\right)\right],
\end{equation}
where $\alpha$ is the strength of the Yukawa deviation with respect to gravity and $\lambda$ its range. $G$ is a constant that matches Newton's gravitational constant, as it would be measured in a Cavendish experiment in the limit $r\gg\lambda$.

\begin{figure}
\begin{center}
\includegraphics[width=0.6\textwidth]{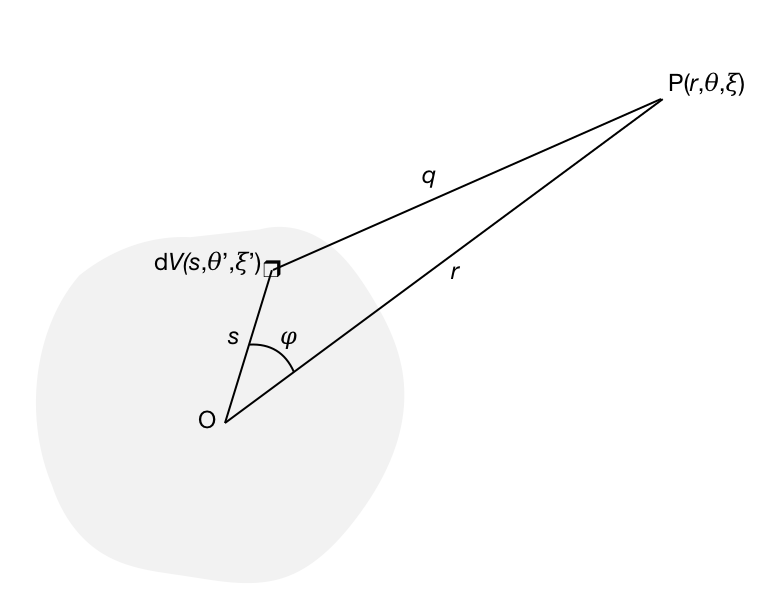}
\end{center}
\caption{Geometry of the problem. We compute the gravitational potential at point $P$ due to a  source (grey area) whose centre-of-mass is  $O$.  In a spherical coordinates system centreed on $O$, $\theta$ (resp. $\theta'$) is the co-latitude of $P$  (resp. of the infinitesimal volume that sources the field at $P'$) and $\xi$ (resp. $\xi'$) its longitude.We define the orthonormal basis $({\bf u}_r, {\bf u}_\theta, {\bf u}_\xi)$ in such a way that ${\bf OP}= r {\bf u}_r$ and ${\bf OP'}= s{\bf u}_{r'}$ so that $\cos\varphi={\bf u}_{r}.{\bf u}_{r'}$.}
\label{fig_geometry}
\end{figure}

It follows that the gravitational potential generated by an extended source is obtained by integrating Eq. (\ref{eq_upointmass}) over the source
\begin{equation}\label{e.V}
U({\mathbf r}) = \int_V U_{\rm pm}({\mathbf r} - {\mathbf s}) {\rm d}^3V,
\end{equation}
where ${\mathbf s}$ is the position-vector of the infinitesimal element of volume ${\rm d}^3V$ and ${\mathbf r}=(r,\theta,\xi)$ are the spherical coordinates of  the point $P$ where the potential is evaluated (see Fig. \ref{fig_geometry}), where $\theta$ is the co-latitude, and $\xi$ the longitude.

As usual, we relate the multipolar decomposition of  this potential to that of the source. To that purpose, we use the standard expansion
\begin{equation} \label{eq_1r}
\frac{1}{q} = \frac{1}{r} \sum_{\ell=0}^\infty \left(\frac{s}{r}\right)^\ell P_\ell(\cos\varphi),
\end{equation}
where $q\equiv \vert{\bf r}-\bf{s}\vert$ and $P_\ell$ are Legendre polynomials. $r$ and $s$ are the distances between the centre of mass of the source and, respectively, the point where we compute the gravitational potential or the infinitesimal volume element of the source so that $s/r<1$; see Fig. \ref{fig_geometry} for the definitions. The Yukawa contributions can be expanded in a similar way thanks to (see Ref.~ \cite{abramowitz+stegun})
\begin{equation} \label{eq_as}
\frac{\rm{e}^{-q/\lambda}}{q} = \frac{1}{\sqrt{rs}} \sum_{\ell=0}^{\infty}(2\ell+1) K_{\ell+\frac{1}{2}}\left( \frac{r}{\lambda} \right) I_{\ell+\frac{1}{2}} \left( \frac{s}{\lambda} \right) P_\ell(\cos \varphi),
\end{equation}
where $I_{\ell+\frac{1}{2}}$ and $K_{n+\frac{1}{2}}$ are modified spherical Bessel functions of the second and third kinds.

Inserting the decompositions~(\ref{eq_1r}-\ref{eq_as}) in Eq.~(\ref{e.V}) and expanding the Legendre polynomials in spherical harmonics $Y_{\ell m}$\footnote{Another common normalization for spherical harmonics is $P_\ell(\cos\varphi) = \frac{4\pi}{2\ell+1} \sum_{m=-\ell}^\ell Y^*_{\ell m}(\theta',\xi') Y_{\ell m}(\theta, \xi$): we use Eq. (\ref{eq_pell}) to ensure that in the case of a homogeneous sphere, we recover $y^N_{00}=1$.}  as
\begin{equation} \label{eq_pell}
P_\ell(\cos\varphi) = \frac{1}{2\ell+1} \sum_{m=-\ell}^\ell Y^*_{\ell m}(\theta',\xi') Y_{\ell m}(\theta, \xi),
\end{equation}
the gravitational potential can be expressed as a multipolar decomposition as
\begin{equation} \label{eq_potential}
 U(P)= U(r,\theta,\xi) = -\frac{G M_\Earth}{r} \sum_{\ell=0}^\infty \sum_{m=-\ell}^{\ell} \left( \frac{R_\Earth}{r}\right)^\ell y_{\ell m}(r) Y_{\ell m}(\theta,\xi),
\end{equation}
where $M_\Earth$ and  $R_\Earth$ are the mass  and equatorial radius of the Earth. We shall use equivalently the notations $Y_{\ell m}(\theta,\xi)$ or $Y_{\ell m}({\bf u}_r)$ in the following.

The introduction  of a Yukawa interaction does not modify the general multipolar expansion of the Earth gravitational potential. Indeed, the multipolar coefficients $y_{\ell m}$ now enjoy two contributions and split as
\begin{equation}
 y_{\ell m}(r) = y^N_{\ell m} + y^Y_{\ell m}(r)
\end{equation}
where the superscripts $N$ and $Y$ stand for the Newton and Yukawa contributions. After trivial algebra, one gets that
\begin{equation} \label{eq_ylmN}
y^N_{\ell m} = \frac{1}{(2\ell+1)M_\Earth} \int_V {\rm d}^3V \rho(s {\bf u}_{r'}) \left( \frac{s}{R_\Earth}\right)^\ell Y^*_{\ell m}({\bf u}_{r'})
\end{equation}
and
\begin{equation} \label{eq_ylmY}
y^Y_{\ell m}(r) = \frac{\alpha}{M_\Earth} \left( \frac{r}{\lambda}\right)^{\ell+\frac{1}{2}} K_{\ell+\frac{1}{2}}\left( \frac{r}{\lambda}\right)
\int_V {\rm d}^3V \rho(s {\bf u}_{r'}) \left( \frac{s}{R_\Earth}\right)^\ell \left( \frac{\lambda}{s} \right)^{\ell+\frac{1}{2}} I_{\ell+\frac{1}{2}} \left( \frac{s}{\lambda} \right) Y^*_{\ell m}({\bf u}_{r'})
\end{equation}
where $\rho(s {\bf u}_{r'})$ is the Earth's density in $P'$ and ${\rm d}^3V=s {\rm d}s\, {\rm d}^2{\bf u}_{r'}$ is the volume element around $P'$. The Earth density can then be expanded in spherical harmonics as
\begin{equation}\label{e.rho}
 \rho(s {\bf u}_{r'}) = \sum_{\ell m} \rho_{\ell m}(s)Y_{\ell m}({\bf u}_{r'}),
\end{equation}
so that we finally get, after integrating over ${\rm d}^2{\bf u}_{r'}$,
\begin{equation}\label{e.ylmr}
y_{\ell m}(r) = \frac{1}{(2\ell+1)M_\Earth} \int s^2 \left(\frac{s}{R_\Earth}\right)^\ell \rho_{\ell m}(s) \left[ 1 + \alpha {\mathcal A}_l\left(\frac{s}{\lambda}\right) {\mathcal B}_l\left(\frac{r}{\lambda}\right) \right] {\rm d}s,
\end{equation}
with the two functions
\begin{eqnarray}
 {\cal A}_\ell(x) &=& x^{-(\ell+1/2)}I_{\ell+1/2}(x)\\
  {\cal B}_\ell(x) &=& (2\ell+1)x^{\ell+1/2} K_{\ell+1/2}(x).
\end{eqnarray}
As expected, the kernel is $m$-independent so that the $m$-dependence arises only from the one of the density. Note that in Eq. (\ref{e.ylmr}) the integral is 1-dimensional. Indeed  $s$ is defined by the Earth surface $R_\Earth({\bf u}_{r'})$ so  is directionally dependent. Since we have performed a multipolar expansion, we need to take this boundary conditions into account in the function $\rho$ so that
\begin{equation}
 \rho(s {\bf u}_{r'}) =  \rho_\Earth(s {\bf u}_{r'})\left\lbrace 1 - \Theta[s-R_\Earth({\bf u}_{r'})] \right\rbrace
\end{equation}
where $\Theta$ is the Heaviside function
The shape of the Earth is thus contained in the multipoles $\rho_{\ell m}$.

\subsection{Gravitational acceleration}

The gravitational acceleration is defined, as usual, as
\begin{equation}
 {\bf g}(r,\theta,\xi) = -{\bf\nabla} U(r,\theta,\xi).
\end{equation}
We thus need to evaluate the gradient of Eq. (\ref{eq_potential}) in spherical coordinates. It decomposes on the spherical basis as
\begin{equation}
  {\bf g}= {\bf g}_\parallel + {\bf g}_\perp,
  \qquad {\rm with}\qquad
  {\bf g}_\parallel \equiv g_r\,{\bf u}_r, \quad
   {\bf g}_\perp \equiv g_\theta\, {\bf u}_\theta + g_\xi \,{\bf u}_\xi.
\end{equation}

\subsubsection{Radial component}

The derivation with respect to $r$ does not act on the spherical harmonics so that
\begin{equation}
g_r (r{\bf u}_r) \equiv \sum_{\ell m} g_r^{\ell m}(r)\, Y_{\ell m}({\bf u}_r)
\end{equation}
with
\begin{equation}\label{grtemp}
g_r^{\ell m}(r)= -\frac{G M_\Earth}{r^2}\left(\frac{R}{r}\right)^\ell \left[ (\ell+1)\left(y^N_{\ell m} +  y^Y_{\ell m}\right)-\frac{r}{\lambda}\dot y^Y_{\ell m} \right]
\end{equation}
where a dot refers to a derivative with respect to $x=r/\lambda$. Since
\begin{eqnarray}
 (\ell+1)y^Y_{\ell m}-\frac{r}{\lambda}\dot y^Y_{\ell m} &=& y^Y_{\ell m}\left[\ell+1+\frac{r}{\lambda}\frac{K_{\ell-1/2}(r/\lambda) }{K_{\ell+1/2}(r/\lambda) }\right].
\end{eqnarray}
Then, it is clear that Eq.~(\ref{grtemp}) recasts as
\begin{equation} \label{eq_grlm}
g^{\ell m}_r= -\frac{GM_\Earth}{r^2} (\ell+1) \left(\frac{R_\Earth}{r}\right)^\ell z_{\ell m}(r)
\end{equation}
where $z_{\ell m}$ is a radial function that differs from $y_{\ell m}$ only by its Kernel,
\begin{equation} \label{eq_zrlm}
z_{\ell m}(r)=  \frac{1}{(2\ell +1)M_\Earth} \int s^2 \left(\frac{s}{R_\Earth}\right)^\ell \rho_{\ell m}(s)
\left[1+\alpha {\cal A}_\ell\left(\frac{s}{\lambda}\right)  {\cal C}_\ell\left(\frac{r}{\lambda}\right) \right] {\rm d}s
\end{equation}
where we have introduced the function
\begin{equation}
 {\cal C}_\ell(x) = (2\ell+1)x^{\ell+1/2} K_{\ell+1/2}(x)\left[1+\frac{x}{\ell+1}\frac{K_{\ell-1/2}(x) }{K_{\ell+1/2}(x) }\right].
\end{equation}

\subsubsection{Angular part}

The angular components are  given by
\begin{eqnarray}
g_\theta(r {\bf u}_r) &  =& \frac{G M_\Earth}{r^2}\sum_{\ell m} \left(\frac{R_\Earth}{r}\right)^\ell y_{\ell m}(r)\partial_\theta Y_{\ell m}({\bf u}_r)\\
g_\xi(r {\bf u}_r) & =& \frac{G M_\Earth}{r^2}\sum_{\ell m} \left(\frac{R_\Earth}{r}\right)^\ell y_{\ell m}(r)\frac{1}{\sin\theta}\partial_\xi Y_{\ell m}({\bf u}_r),
\end{eqnarray}
with $y_{\ell m}(r)$ given by Eq.~(\ref{e.ylmr}). However such a decomposition does not give a proper multipolar expansion since $\partial_\theta Y_{\ell m}$ mixes different multipoles. Indeed, after derivation the expansion is no more in an orthonormal basis. The standard way to express the gravitational acceleration in a good frame is to use recursion properties between spherical harmonics (see e.g. Refs.~\cite{cunningham70,metris98,fantino09,petrovskaya10}). Here, we propose to use a simpler way by introducing spin-weighted spherical harmonics.

To that purpose, we first define the two complex vectors
\begin{equation}
 {\bf u}_\pm \equiv \frac{1}{\sqrt{2}}\left(   {\bf u}_\theta\mp j {\bf u}_\xi \right),
\end{equation}
where $j^2=-1$ so that
\begin{equation}
{\bf g}_\perp=  g_+ {\bf u}_+ +  g_-  {\bf u}_-
\qquad{\rm with}\qquad
g_\pm = \frac{1}{\sqrt{2} }\left( g_\theta \pm j g_\xi \right).
\end{equation}

With our notations, we get
\begin{equation}
g_\pm = \frac{1}{\sqrt{2}} \frac{GM_\Earth}{r^2}\sum_{\ell m}\left(\frac{R_\Earth}{r}\right)^\ell y_{\ell m}(r)
\left[\partial_\theta \pm \frac{j}{\sin\theta}\partial_\xi  \right] Y_{\ell m}({\bf u}_r).
\end{equation}
Now acting  $s$-times with the complex derivative in the square brackets on the spherical harmonics defines the spin-weighted spherical harmonics \cite{newman66, goldberg67}
\begin{eqnarray}
{}_s Y_{\ell m}(\theta,\xi) &=& \sqrt{\frac{2\ell +1}{4\pi}} {\cal D}_{-s m}^\ell (\theta,\xi,0)
\end{eqnarray}
where ${\cal D}_{-s m}^\ell$ stands for the Wigner matrices. More explicitly,
\begin{eqnarray}
{}_s Y_{\ell m}(\theta,\xi)  &=& \sqrt{\frac{2\ell +1}{4\pi}\frac{(\ell+m)!(\ell-m)!}{(\ell+s)!(\ell-s)!}} \left(\sin\theta/2\right)^{2\ell} \sum_r
   \left(\begin{array}{c}\ell-s \\ r   \end{array} \right)   \left(\begin{array}{c}\ell+s \\ r+s -m   \end{array} \right)\nonumber\\
   && \times (-1)^{\ell-r-s }\hbox{e}^{jm\xi}\left({\rm cot}\theta/2\right)^{2r+s-m},
\end{eqnarray}
where in the ${}_s Y_{\ell m}$, $s$ is an integer which obviously does not refer to the radial coordinate. These spin-weighted spherical harmonics form an orthonormal basis, i.e. they satisfy
$$
\int {\rm d}^2{\bf n} \,\,{}_s Y_{\ell m}^* ({\bf n})\,{}_s Y_{\ell' m'}({\bf n})=\delta_{\ell\ell'}\delta_{mm'}.
$$
From Eq.~(2.7) of Ref.~\cite{goldberg67}), we have
\begin{eqnarray}
\left[\partial_\theta \pm \frac{j}{\sin \theta} \partial \xi\right] Y_{\ell m}(\theta,\xi) &=& \pm\sqrt{\ell(\ell +1)} {}_{\pm1}Y_{\ell m}(\theta,\xi).
\end{eqnarray}
from which it follows that the proper multipolar expansion of the gravitational acceleration is
\begin{equation} 
g_\pm =  \sum_{\ell m}g_\pm^{\ell m} {}_{\pm1}Y_{\ell m}(\theta,\xi)
\end{equation}
with
\begin{equation} \label{eq_gpmlm}
g_\pm^{\ell m} =  \pm \frac{1}{\sqrt{2}}\frac{GM_\Earth}{r^2}\sqrt{\ell(\ell+1)}\left(\frac{R_\Earth}{r}\right)^\ell y_{\ell m}(r).
\end{equation}

\subsubsection{Summary}

Equations (\ref{eq_grlm}) for the radial component and (\ref{eq_gpmlm}) for the angular component allow us to  compute directly the contribution of the ($\ell$, $m$) multipole to the gravitational acceleration of any extended body once $\rho(s {\bf u}_{r'})$ is known. They now need to be translated to non-Keplerian perturbations applied to bodies orbiting around the Earth.

\subsection{Orbital perturbations and secular effect on satellites osculating parameters}

Given perturbing forces acting on a satellite, the Lagrange-Gauss equations allow one to compute the secular variations of the satellite's orbital parameters~\cite{kaula66, uzan, roy05}. They can be also used to deduce the characteristics of a perturbing source from a measurement of the satellite's dynamics. In particular, they can be used to estimate the Earth gravitational field spherical harmonic coefficients.
This section establishes the Lagrange-Gauss equations and the expression of the perturbing force arising from the shape of the Earth and a non-Newtonian gravity modelled by a Yukawa potential.

\subsubsection{Expression of the perturbing force}

\begin{figure}
\begin{center}
\includegraphics[width=.7\textwidth]{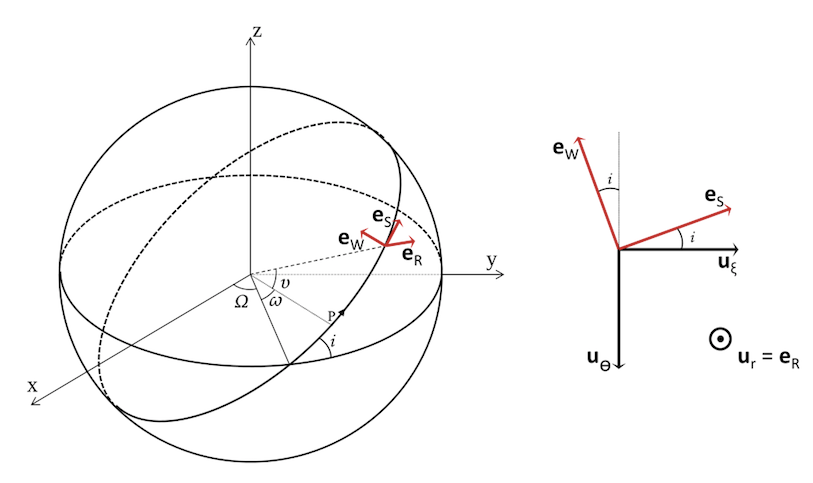}
\caption{Left: Comoving frame associated to the orbit and in which we decompose the perturbation. Notice that we have introduced $i$ the inclination, $\Omega$ the longitude of the ascending node, $\omega$ the argument of the perigee $P$  and $\nu$ the true anomaly.   Right: A rotation to transform the spherical coordinates unit vectors ${\bf u}_\theta$ and ${\bf u}_\xi$ into the comoving unit vectors ${\bf e}_S$ and ${\bf e}_W$.}
\label{fig_coords}
\end{center}
\end{figure}

Let us define a perturbing force acting on an orbiting body as the difference between the  actual force applied to the body and the pure Newtonian monopole gravitational force. We ignore all non-gravitational forces, as well as gravity from the Sun, the Moon and other Solar system planets, so that the perturbing force is
\begin{equation}
{\bf F}_{\rm pert}={\bf g} + \frac{GM_\Earth}{r^2} {\bf u}_r = \mathcal{R} {\bf e}_R + \mathcal{S} {\bf e}_S + \mathcal{W} {\bf e}_W
\end{equation}
 where $\mathcal{R}$, $\mathcal{S}$ and $\mathcal{W}$ are the radial, tangential (in the orbital plane) and orthogonal components of the perturbing force per unit of reduced mass. The unit vectors ${\bf e}_R$, ${\bf e}_S$, ${\bf e}_W$ are defined in Fig.~\ref{fig_coords}. Since ${\bf e}_R={\bf u}_r$ the change of coordinates between the two frames reduce to a rotation  so that the components of the force are
 \begin{eqnarray} \label{eq_rsw1}
\mathcal{R} & = & \frac{m_{\rm sat}}{\mu}\left[ g_r(r,\theta,\xi) + \frac{GM_\Earth}{r^2}\right] \\
 \label{eq_rsw2}
\mathcal{S} & = & -\frac{m_{\rm sat}}{\mu} \left[g_\theta(r,\theta,\xi) \sin i -  g_\xi(r,\theta,\xi) \cos i \right] \\
 \label{eq_rsw3}
\mathcal{W} & = & -\frac{m_{\rm sat}}{\mu} \left[ g_\theta(r,\theta,\xi) \cos i +  g_\xi(r,\theta,\xi) \sin i\right]
\end{eqnarray}
where $\mu = M_\Earth m_{\rm sat} / (M_\Earth + m_{\rm sat})$ is the reduced mass of the Earth-satellite system.

The perturbation force can also be decomposed in the basis $({\bf u}_r,{\bf u}_+,{\bf u}_-)$ as
\begin{equation}
{\bf F}_{\rm pert}= {\cal R}{\bf u}_r + \sum_{s=\pm} {\cal R}^s {\bf u}_s
\end{equation}
so that its angular component can be decomposed in spin-weighted harmonics as
 \begin{equation}
 {\cal R}^\pm = \frac{1}{\sqrt{2} }\left({\cal S} \pm j {\cal W} \right) = \mp \frac{m_{\rm sat}}{\mu} \exp\left[j\left(\frac{\pi}{2}\mp i\right)\right] g_\pm.
\end{equation}

It follows that the multipolar decomposition is now well-defined as
\begin{equation}\label{e.fperp}
{\bf F}_{\rm pert}= \sum_{\ell m}\left[{\cal R}_{\ell m} Y_{\ell m} {\bf u}_r + \sum_{s=\pm} {\cal R}^s_{\ell m} \, {}_s Y_{\ell m}{\bf u}_s
\right]
\end{equation}
with
\begin{eqnarray}
{\cal R}_{\ell m} &=& \frac{G M_{\rm tot}}{r^2} \left[ - (\ell+1) \left(\frac{R_\Earth}{r}\right)^\ell z_{\ell m}(r) + \delta_{\ell 0}\delta_{m0}  \right], \\
 {\cal R}_{\ell m}^\pm &=& \mp \frac{1}{\sqrt{2}} \frac{G M_{\rm tot}}{r^2}
 \sqrt{\ell(\ell+1)}\left(\frac{R_\Earth}{r}\right)^\ell  y_{\ell m}(r)
 \exp\left[j\left(\frac{\pi}{2}\mp i\right)\right],
\end{eqnarray}
where $M_{\rm tot} = M_\Earth+m_{\rm sat}$ and $y_{\ell m}(r)$  and $z_{\ell m}(r)$ are defined in Eqs.~(\ref{e.ylmr}) and~(\ref{eq_zrlm}) respectively.

It immediately follows that the radial, tangential and othogonal components of the perturbing force (Eqs. \ref{eq_rsw1}-\ref{eq_rsw3}) are
\begin{equation} \label{eq_rsw21}
{\mathcal R}(r,\theta,\xi) = \frac{GM_{\rm tot}}{r^2}\sum_{\ell m} \left[ - (\ell+1) \left(\frac{R_\Earth}{r}\right)^\ell z_{\ell m}(r) + \delta_{\ell 0}\delta_{m0}  \right] Y_{\ell m}(\theta,\xi)
\end{equation}
\begin{equation} \label{eq_rsw22}
{\mathcal S}(r,\theta,\xi) = -\frac{j}{2} \frac{GM_{\rm tot}}{r^2}\sum_{\ell m} \sqrt{\ell(\ell+1)} \left(\frac{R_\Earth}{r}\right)^\ell y_{\ell m}(r) \left[ {\rm e}^{-ji} {}_{+1}Y_{\ell m}(\theta,\xi) + {\rm e}^{ji} {}_{-1}Y_{\ell m}(\theta,\xi)\right]
\end{equation}
\begin{equation} \label{eq_rsw23}
{\mathcal W}(r,\theta,\xi) = - \frac{1}{2}\frac{GM_{\rm tot}}{r^2}\sum_{\ell m} \sqrt{\ell(\ell+1)} \left(\frac{R_\Earth}{r}\right)^\ell y_{\ell m}(r) \left[ {\rm e}^{-ji} {}_{+1}Y_{\ell m}(\theta,\xi) - {\rm e}^{ji} {}_{-1}Y_{\ell m}(\theta,\xi)\right]
\end{equation}

In Newtonian gravity, the perturbations arise only from the shape of the Earth, from the gravitational perturbations of other celestial bodies (Sun, Moon, planets, etc.) and of friction forces from the atmosphere and radiation pressure. They all have been studied in details and shown to cause secular drifts such as the precession of the line of nodes (the latter being mostly sourced by the Earth equatorial bulge through the $J_2 \equiv -\sqrt{5} y_{20}$ zonal term)~\cite{roy05}. In a theory gravity beyond Einstein (or, on small scales, Newton), the existence of an extra gravitational potential causes a new set of perturbations, also related to the shape of the Earth. In our model at hand, the Yukawa potential mixes the shape of the Earth contributions and the non-Newtonian interaction.

\subsubsection{Lagrange-Gauss equations and secular effects}

Secular effects due to a perturbing force on osculating parameters for a given orbit configuration can be computed from the Lagrange-Gauss equations once the component of the perturbative force~(\ref{e.fperp}) are known.

The Lagrange-Gauss equations then read
\begin{equation} \label{e.gaussdebut}
\frac{{\rm d}a}{{\rm d}t} = 2\sqrt{\frac{a^3}{GM_{\rm tot}(1-e^2)}} \left[\mathcal{R} e \sin\nu + (1+e\cos\nu) \mathcal{S}\right] 
\end{equation}
\begin{equation}
\frac{{\rm d}e}{{\rm d}t} = \sqrt{\frac{a(1-e^2)}{GM_{\rm tot}}} \left[ \mathcal{R}\sin\nu + \frac{e+2\cos\nu + e\cos^2\nu}{1+e\cos\nu} \mathcal{S} \right]
\end{equation}
\begin{equation}
\frac{{\rm d}i}{{\rm d}t} = \sqrt{\frac{a(1-e^2)}{GM_{\rm tot}}} \frac{\cos(\omega+\nu)}{1+e\cos\nu} \mathcal{W} \\
\frac{{\rm d}\Omega}{{\rm d}t} = \sqrt{\frac{a(1-e^2)}{GM_{\rm tot}}} \frac{\sin(\omega+\nu)}{1+e\cos\nu} \frac{\mathcal{W}}{\sin i}
\end{equation}
\begin{equation}
\frac{{\rm d}\omega}{{\rm d}t} = \sqrt{\frac{a(1-e^2)}{GM_{\rm tot}}} \left[ -\frac{\mathcal{R}}{e} \cos\nu + \frac{(2+e\cos\nu)\sin\nu}{e(1+e\cos\nu)} \mathcal{S} - \frac{\sin(\omega+\nu)}{1+e\cos\nu} \frac{\cos i}{\sin i} \mathcal{W} \right]
\end{equation}
\begin{equation} \label{e.gaussfin}
\frac{{\rm d}\ell}{{\rm d}t} = n + \sqrt{\frac{a}{GM_{\rm tot}}} \frac{1-e^2}{e(1+e\cos\nu)} \left[ \mathcal{R} (-2e+\cos\nu+2\cos^2\nu) - \sin\nu (2+e\cos\nu) \mathcal{S} \right] 
\end{equation}
where $a$ is the semi-major axis of the orbit, $e$ its eccentricity, $i$ its inclination, $\Omega$ the longitude of the ascending node, $\omega$ the argument of the perigee and $\ell = n(t-T)$ \footnote{Note that this $\ell$ is not to be confused with the multipole of the spherical harmonics expansion. We will not used it in the remainder of this paper} with $n \equiv \sqrt{\mu/a^3}$.  Additionally, the true anomaly variation is given by
\begin{equation}
\frac{{\rm d}\nu}{{\rm d}t} = \sqrt{\frac{\mu}{a^3(1-e^2)^3}}(1+e \cos\nu)^2,
\end{equation}
and does not depend on the perturbing force. The $\Omega$, $\omega$, $i$ and $\nu$ angles are shown in the left panel of Fig. \ref{fig_coords}.

\subsubsection{Summary}

This provides all the equations for studying the orbital motion of a satellite in a theory of gravity including a Yukawa interaction together with the Newtonian force. The Lagrange-Gauss equations~(\ref{e.gaussdebut}-\ref{e.gaussfin}) can be solved for the secular effects caused by a Yukawa interaction on satellites dynamics, once the perturbing force~(\ref{e.fperp}) is known. The latter depends on the shape and mass distribution of the Earth, given by Eq.~(\ref{e.rho}) through the $y_{\ell m}(r)$ functions, explicitly given in Eq.~(\ref{e.ylmr}). As we have already emphasised, the parameter $\rho_{\ell m}$ describing the Earth and $(\alpha,\lambda)$ describing the non-Newtonian gravity are entangled. This formalism paves the way to many investigations to which we now turn.

\section{Worked-out example: $N$-layer, rotationally symmetric Earth model} \label{sect_emodel}

This section applies our previous formalism to a simple model of the Earth. It will  allow us to better grasp the impact of the Yukawa interaction on the Earth gravitational field and the way it mixes with the usual perturbing effects arising from the shape of the Earth.

To that purpose,  we consider a $N$-layer Earth, where each layer, of radius $R_i(\theta,\xi)$, is homogeneous with density $\rho_i$, such that
\begin{equation}
\rho(s {\bf u}_{r'}) = \sum_{i=1}^N \rho_i \lbrace\Theta[s - R_{i-1}({\bf u}_{r'})] - \Theta[s - R_i({\bf u}_{r'})]\rbrace,
\end{equation}
where $R_0({\bf u}_{r'})=0$.

\subsection{Monopole and quadrupole} \label{ssect_y00y20}

The $y_{\ell m}$ coefficients are defined in Eq.~(\ref{e.ylmr}) and have two components given in  Eqs. (\ref{eq_ylmN}) and (\ref{eq_ylmY}). Introducing the term $q^Y_{\ell m}$ such that
$$
y^Y_{\ell m}(r) \equiv \frac{\alpha}{M_\Earth} \left( \frac{r}{\lambda}\right)^{\ell+\frac{1}{2}} K_{\ell+\frac{1}{2}}\left( \frac{r}{\lambda}\right) q^Y_{\ell m}
$$
and making explicit the volume integral in spherical coordinates (we detail the computation only for $y^Y_{\ell m}$ since the one of $y^N_{\ell m}$ derives trivially from it), we write
\begin{equation}
q^Y_{\ell m} = \int_0^{2\pi}{\rm d}\xi \int_0^\pi {\rm d}\theta \sin\theta Y^*_{\ell m}(\theta,\xi) Q^Y_\ell(\theta,\xi)
\end{equation}
with
\begin{equation} \label{eq_Qn}
Q^Y_\ell(\theta,\xi) \equiv \int_0^{R(\theta,\xi)} {\rm d}s \rho(s, \theta, \xi) \frac{s^{\ell+2}}{R^\ell_\Earth} \left( \frac{\lambda}{s} \right)^{\ell+\frac{1}{2}} I_{\ell+\frac{1}{2}} \left( \frac{s}{\lambda} \right).
\end{equation}
Then, we introduce the function
\begin{equation} \label{eq_phin}
\phi_\ell(x, k) = 2^{-\ell-\frac{3}{2}} x^{\ell+3} \frac{\Gamma\left( \frac{\ell+3}{2}\right)}{\Gamma\left( \ell+\frac{3}{2} \right) \Gamma\left(\frac{\ell+5}{2} \right)} {}_1F_2\left(\frac{\ell+3}{2}; \ell+\frac{3}{2}, \frac{\ell+5}{2}; \frac{k^2x^2}{4} \right),
\end{equation}
where $\Gamma$ is the Gamma function and ${}_1F_2$ is a generalised hypergeometric function. We note that (see~\ref{app_phi} for an explicit proof)
\begin{equation} \label{eq_H}
\int_a^b {\rm d}x x^{\ell+2} \left( \frac{\lambda}{R_\Earth x}\right)^{\ell+\frac{1}{2}} I_{\ell+\frac{1}{2}} \left( \frac{R_\Earth x}{\lambda}\right) = \phi_\ell \left(b, \frac{R_\Earth}{\lambda} \right) - \phi_\ell \left(a, \frac{R_\Earth}{\lambda} \right),
\end{equation}
and letting $x=s/R_\Earth$, Eq. (\ref{eq_Qn}) becomes
\begin{equation} \label{eq_QRRc}
Q_\ell^Y(\theta,\xi) = R_\Earth^3 \sum_{i=1}^N \left[ \phi_\ell\left( \frac{R_i(\theta,\xi)}{R_\Earth}, \frac{R_\Earth}{\lambda}\right) - \phi_\ell\left(\frac{R_{i-1}(\theta,\xi)}{R_\Earth}, \frac{R_\Earth}{\lambda} \right) \right].
\end{equation}

We now further assume that the Earth is made of $N$ concentric, homogeneous ellipsoidal, rotationally symmetric layers. Noting $f_i=(R_{{\rm eq}, i} - R_{{\rm pole}, i})/R_{{\rm eq}, i}$ the $i$th layer's flattening, where $R_{{\rm eq}, i}$ and $R_{{\rm pole}, i}$ are its equatorial and polar radiuses, we get
\begin{equation} \label{eq_geoid}
R_i(\theta,\xi) = \frac{R_{{\rm eq}, i} (1-f_i)}{\sqrt{1-(2f_i-f_i^2)\sin^2\theta}}.
\end{equation}

We now explicitly compute the first two non-zero spherical harmonics coefficients (monopole and quadrupole), $y_{00}$ and $y_{20}$, usually called $\overline{C_{00}}$ and $\overline{C_{20}}$ in the literature. Note that with our normalization, $y_{00} = \overline{C_{00}}$ and $y_{20} = \overline{C_{20}}$. 
The quadrupole $y_{20}$ is linked to the $J_2$ flattening of the Earth via $J_2 = -\sqrt{5}y_{20}$ if we ignore the rotation of the Earth.

Under the rotational symmetry assumption, $y_{\ell m}=0$ for all $m\neq0$, and
\begin{multline} \label{eq_yl0n}
y_{\ell 0}^N = \frac{2\pi}{\sqrt{2\ell+1}(\ell+3)M_\Earth} \int_0^\pi {\rm d}\theta\sin\theta P_{\ell 0}(\cos\theta) R_\Earth^3 \\
\times \sum_{i=1}^N \rho_i \left[\left( \frac{R_{{\rm eq}, i}(1-f_i)}{R_\Earth\sqrt{1-(2f_i-f_i^2)\sin^2\theta}}\right)^{\ell+3} - \left( \frac{R_{{\rm eq}, i-1}(1-f_{i-1})}{R_\Earth\sqrt{1-(2f_{i-1}-f_{i-1}^2)\sin^2\theta}}\right)^{\ell+3} \right]
\end{multline}
and
\begin{multline}\label{eq_yl0y}
y_{\ell 0}^Y(r) = \frac{2\pi\alpha}{M_\Earth} \sqrt{2\ell+1} \left( \frac{r}{\lambda}\right)^{\ell+\frac{1}{2}} K_{\ell+\frac{1}{2}}\left( \frac{r}{\lambda}\right) \int_0^\pi {\rm d}\theta \sin\theta P_{\ell 0}(\cos\theta) R_\Earth^3 \\
\times \sum_{i=1}^N \rho_i \left[ \phi_\ell\left( \frac{R_{{\rm eq}, i}(1-f_i)}{R_\Earth\sqrt{1-(2f_i-f_i^2)\sin^2\theta}}, \frac{R_\Earth}{\lambda} \right) -
\phi_\ell\left( \frac{R_{{\rm eq}, i-1}(1-f_{i-1})}{R_\Earth\sqrt{1-(2f_{i-1}-f_{i-1}^2)\sin^2\theta}}, \frac{R_\Earth}{\lambda} \right) \right]
\end{multline}
where $P_{\ell 0}(\cos\theta)$ is an associated Legendre polynomial.

The computation involves integrating hypergeometric functions over $\theta$, which can easily be done numerically, but requires further assumptions to allow for an analytic expression. Assuming that the flattening of the Earth layers are small ($f_i \ll 1$), we can Taylor expand the $\phi_\ell$ functions, and we obtain the monopole (at first order in $f_i$)

\begin{multline} \label{eq_y00}
y_{00}(r) = \frac{4\pi  R_\Earth^3}{3M_\Earth}\sum_{i=1}^N \rho_i \left[ (1-f_i) \frac{R_{{\rm eq}, i}^3}{R_\Earth^3} - (1-f_{i-1})\frac{R_{{\rm eq}, i-1}^3}{R_\Earth^3}  \right] \\
+ \frac{4\pi\alpha}{3M_\Earth} {\rm e}^{-r/\lambda} R_\Earth^3 \sum_{i=1}^2 \rho_i \left\{ 3 \frac{\lambda^3}{R_\Earth^3} \left[ \frac{R_{{\rm eq}, i}}{\lambda} \cosh \frac{R_{{\rm eq},i}}{\lambda} - \left( \frac{f_i R_{{\rm eq}, i}^2}{3\lambda^2} + 1 \right) \sinh \frac{R_{{\rm eq}, i}}{\lambda} \right] \right. \\
\left. - 3 \frac{\lambda^3}{R_\Earth^3} \left[ \frac{R_{{\rm eq}, i-1}}{\lambda} \cosh \frac{R_{{\rm eq},i-1}}{\lambda} - \left( \frac{f_{i-1} R_{{\rm eq}, i-1}^2}{3\lambda^2} + 1 \right) \sinh \frac{R_{{\rm eq}, i-1}}{\lambda} \right] \right\},
\end{multline}
where the first term is the Newtonian contribution (also computed under the assumption $f_i \ll 1$).

In the simple case of a homogeneous ellipsoid ($N=1$), Eq. (\ref{eq_y00}) reduces to
\begin{equation}
y_{00}(r) = 1+\frac{\alpha}{1-f} \Phi\left(\frac{R_\Earth}{\lambda},f\right) {\rm e}^{-r/\lambda} ,
\end{equation}
where we used that, for a homogeneous ellipsoid
\begin{equation} \label{eq_fell}
M_\Earth = \frac{4}{3}\pi R_\Earth^3 \rho (1-f),
\end{equation}
and where
\begin{equation}
\Phi(x,f) = 3 \frac{x \cosh(x) - \sinh(x)}{x^3} - \frac{\sinh x}{x} f
\end{equation}
generalises the usual sphere's form factor \cite{adelberger03} to an ellipsoid of flatness $f$. In the case of a homogeneous sphere, we thus recover the result from the direct integration over the sphere \cite{adelberger03}, and in the case of a two-layer spherical Earth, we recover the expression given in Ref.~\cite{berge18}. The $\Phi$ function is discussed in \ref{app_phi2}.

Under the same assumption, at first order in $f_i$, we find the quadrupole
\begin{multline} \label{eq_y20}
y_{20}(r) = -\frac{8\pi R_\Earth^3}{15 \sqrt{5} M_\Earth} \sum_{i=1}^N \rho_i \left[ \frac{R_{{\rm eq},i}^5}{R_\Earth^5} f_i - \frac{R_{{\rm eq},i-1}^5}{R_\Earth^5} f_{i-1} \right]   
+ \frac{8\pi\alpha R_\Earth^3}{3\sqrt{5}M_\Earth} {\rm e}^{-r/\lambda} \left( 3+ 3\frac{r}{\lambda} + \frac{r^2}{\lambda^2} \right) \\
\times \sum_{i=1}^N \rho_i \left\{ 3 f_i \frac{\lambda^5}{R_\Earth^5} \left[ \frac{R_{{\rm eq},i}}{\lambda} \cosh \frac{R_{{\rm eq},i}}{\lambda} - \left( \frac{R_{{\rm eq},i}^2}{3\lambda^2} + 1 \right) \sinh \frac{R_{{\rm eq},i}}{\lambda} \right] \right. \\
\left. - 3 f_{i-1} \frac{\lambda^5}{R_\Earth^5} \left[ \frac{R_{{\rm eq},i-1}}{\lambda} \cosh \frac{R_{{\rm eq},i-1}}{\lambda} - \left( \frac{R_{{\rm eq},i-1}^2}{3\lambda^2} + 1 \right) \sinh \frac{R_{{\rm eq},i-1}}{\lambda} \right] \right\}
\end{multline}
where the first term is the Newtonian contribution.

For a homogeneous Earth of density $\rho$ and flattening $f$, Eq. (\ref{eq_y20}) simplifies to
\begin{equation} \label{eq_y20_homogeneous}
y_{20}(r) = -\frac{2f}{5\sqrt{5}(1-f)} \left[ 1
- 5\alpha {\rm e}^{-r/\lambda} \kappa\left(\frac{r}{\lambda}\right) \Phi_2\left( \frac{R_\Earth}{\lambda} \right) \right],
\end{equation}
where we used Eq. (\ref{eq_fell}), $\kappa(x) = 3+3x+x^2$ and where the function
\begin{equation}
\Phi_2(x) = 3 \frac{x \cosh(x) - \left(x^2/3+1\right)\sinh(x)}{x^5}
\end{equation}
is a form factor akin to the $\Phi$ function above (see \ref{app_phi2}).

Eq. (\ref{e.ylmr}), Eqs. (\ref{eq_y00}) and (\ref{eq_y20}) make the role of the Yukawa term clearer. The coefficients of the potential's spherical harmonic expansion obviously depend on where they are estimated, through the exponential decrease of the Yukawa interaction with respect to distance. Another major impact is the presence of a form factor, which emerges because  Gauss' theorem does not apply to a Yukawa interaction (even for a spherical Earth). In other words, it quantifies the fact that for a short-range Yukawa interaction, regions of the Earth close to the experiment play a bigger part in the gravitational field than regions further away, so that the Yukawa interaction created by an extended body is not equal to the Yukawa interaction created by a point-mass of the same mass as the body's.
Hence, measurements of the gravitational field made on the ground, in low earth orbit, or at a greater distance of Earth, will provide different coefficients; we then should be careful when combining different gravity measurements.

In particular, the $y_{00}$ coefficient is not equal to 1 by definition (as in pure Newtonian gravity), but is affected by a supplementary, distance-dependent term $y_{00}(r) = 1 + y_{00}^Y(r)$.
Therefore, it should not be set a priori to 1 when measuring the Earth potential, but estimated like other coefficients. Actually, estimating it is akin to estimating an effective Newton constant that depends on the distance to the centre of the Earth, with the ``real'' Newton constant being estimated by Cavendish-like experiments on the ground.

Similar conclusions can be drawn for the $y_{20}$ term. It is affected by the Yukawa term, whose value will depend on $\alpha$ and on the ratio between the Yukawa range and the characteristic scales of the experiment ($r$ and $R_\Earth$), with a maximum effect around $\lambda \sim r$.
As shown in \ref{app_phi2}, $\Phi_2(R_\Earth/\lambda)$ is of order a few percent in this regime, so that the Yukawa contribution to the $y_{20}$ measured by a satellite orbiting the Earth at a low altitude amounts to a few percent of $\alpha$.

We should finally note that under the assumptions used to obtain Eqs. (\ref{eq_y00}) and (\ref{eq_y20}), higher terms ($y_{40}$, $y_{60}$...) vanish. We need to Taylor expand to higher orders in $f_i$ to get non-zero coefficients. We checked that the approximations (\ref{eq_y00})-(\ref{eq_y20}) provide accurate numbers (up to the percent accuracy) by comparing them to the numerical integration of Eq. (\ref{eq_Qn}) and the corresponding equation for the Newtonian part.

\subsection{Gravitational acceleration} \label{ssect_exg}

The expressions above for the first spherical harmonic coefficients can be inserted in Eqs. (\ref{eq_grlm}) and (\ref{eq_gpmlm}) to derive the expression of the gravitational acceleration of an $N$-layer rotationally symmetric Earth (for which all $m\neq 0$ multipoles cancel).
However, this requires tedious algebra, so we will restrain ourselves to the homogeneous ellipsoid case $N=1$, and consider that only the $\ell=0$ and $\ell=2$ multipoles are non-negligible (this is a reasonable assumption since the measured $J_2$ is 1000 times higher than the following spherical harmonic coefficients).
We find that the norm of the radial and tangential components are
\begin{multline}
g_{||}(r, \theta,\xi) = -\frac{GM_{\rm tot}}{r^2} \left[1+\frac{\alpha}{1-f} \left(1+\frac{r}{\lambda}\right) {\rm e}^{-r/\lambda} \Phi\left(\frac{R_\Earth}{\lambda},f\right)\right] \\
- \frac{3\sqrt{5}}{2} \frac{GM_{\rm tot}}{r^2} z_{20}(r) (3\cos^2\theta-1)
\label{force1}
\end{multline}
where we made $z_{00}(r)$ explicit, and
\begin{equation}
|g_\perp(r, \theta,\xi)| = \sqrt{\frac{5}{2}}\frac{GM_\Earth}{r^2} \left(\frac{R_\Earth}{r}\right)^2 |y_{20}(r)| = \frac{1}{\sqrt{2}}\frac{GM_\Earth}{r^2} \left(\frac{R_\Earth}{r}\right)^2 |J_{20}(r)|
\label{force2}
\end{equation}
where we used that
\begin{equation}
{}_{\pm1}Y_{20}(\theta,\xi)=\mp \frac{1}{4}\sqrt{\frac{15}{2\pi}} \sin2\theta.
\end{equation}

Note that the $J_2$ contribution is formally identical to the Newtonian case, although  now $J_2$ is a function of $r$.
Following the rotational symmetry of our model, those components are indeed independent of the longitude $\xi$.
We provide order-of-magnitude estimates of the Yukawa accelerations and compare them with usual gravitational and non-gravitational perturbations, for a homogeneous Earth, in Sect. \ref{ssect_ofm_pert}.

\subsection{Perturbations and secular effects on satellite dynamics}

The Lagrange-Gauss equations (Eqs. \ref{e.gaussdebut}-\ref{e.gaussfin}) can be trivially obtained for a $N$-layer rotationally symmetric Earth in a way similar to that used to get the gravitational acceleration above, by inserting Eqs. (\ref{eq_yl0n})-(\ref{eq_yl0y}) in Eqs. (\ref{eq_rsw21})-(\ref{eq_rsw23}). 
As for the gravitational acceleration, we restrain ourselves to the homogeneous ellipsoid.
In this case, the components of the perturbing force are
\begin{equation}
{\mathcal R}(r, \theta,\xi) = -\frac{GM_{\rm tot}}{r^2} \left[ \frac{\alpha}{1-f} \left(1+\frac{r}{\lambda}\right) {\rm e}^{-r/\lambda} \Phi\left(\frac{R_\Earth}{\lambda},f\right) + \frac{3\sqrt{5}}{2} z_{20}(r) (3\cos^2\theta-1)\right]
\end{equation}
\begin{equation}
{\mathcal S}(r, \theta,\xi) = \frac{3}{2\sqrt{2\pi}}\frac{GM_{\rm tot}}{r^2} \left(\frac{R_\Earth}{r}\right)^2 J_{20}(r) \sin2\theta \sin i
\end{equation}
\begin{equation}
{\mathcal W}(r, \theta,\xi) = -\frac{3}{2\sqrt{2\pi}}\frac{GM_{\rm tot}}{r^2} \left(\frac{R_\Earth}{r}\right)^2 J_{20}(r) \sin2\theta \cos i
\end{equation}
As was the case for the gravitational acceleration, those components are indeed independent of the longitude $\xi$.

Although we do not solve the Lagrange-Gauss equations in this paper, it is instructive to express the components of the perturbing force as a function of the satellite's unperturbed orbit's Keplerian parameters (which is required to solve the equations).
Using that $r=a(1-e^2)/(1+e\cos\nu)$ and that (following some algebra based on Ref. \cite{montenbruck})
\begin{eqnarray}
\sin2\theta &=& 2\sin(\omega+\nu) \sin i \sqrt{1-\sin^2(\omega+\nu)\sin^2i} \\
\cos\theta &=& \sin(\omega+\nu) \sin i,
\end{eqnarray}
we find that
\begin{multline} \label{eq_rlg}
{\mathcal R} = GM_{\rm tot} \frac{(1+e\cos\nu)^2}{a^2(1-e^2)^2} \left\{ \frac{3}{5} \left(\frac{R_\Earth(1+e\cos\nu)}{a(1-e^2)}\right)^2 \frac{f}{1-f} (3\sin^2(\omega+\nu)\sin^2i - 1) \right. \\
-\frac{\alpha}{1-f} \exp\left(-\frac{a(1-e^2)}{\lambda(1+e\cos\nu)}\right) \left[ \left(1+\frac{a(1-e^2)}{\lambda(1+e\cos\nu)}\right) \Phi\left(\frac{R_\Earth}{\lambda},f\right) \right. \\
\left. \left. + 3\left(\frac{R_\Earth(1+e\cos\nu)}{a(1-e^2)}\right)^2 f \sigma\left(\frac{a(1-e^2)}{\lambda(1+e\cos\nu)}\right) \Phi_2\left(\frac{R_\Earth}{\lambda}\right) (3\sin^2(\omega+\nu)\sin^2i-1)\right]\right\}
\end{multline}
\begin{multline}\label{eq_slg}
{\mathcal S} = \frac{6}{5\sqrt{2\pi}} GM_{\rm tot} R_\Earth^2 \left(\frac{1+e\cos\nu}{a(1-e^2)}\right)^4 \frac{f}{1-f} \sin(\omega+\nu) \sin^2i \sqrt{1-\sin^2(\omega+\nu)\sin^2i} \\
\times \left[1 - 5\alpha \exp\left(-\frac{a(1-e^2)}{\lambda(1+e\cos\nu)}\right) \kappa\left(\frac{a(1-e^2)}{\lambda(1+e\cos\nu)}\right) \Phi_2\left(\frac{R_\Earth}{\lambda}\right) \right]
\end{multline}
\begin{multline}\label{eq_wlg}
{\mathcal W} = -\frac{3}{5\sqrt{2\pi}} GM_{\rm tot} R_\Earth^2 \left(\frac{1+e\cos\nu}{a(1-e^2)}\right)^4 \frac{f}{1-f} \sin(\omega+\nu) \sin(2i) \sqrt{1-\sin^2(\omega+\nu)\sin^2i} \\
\times \left[1 - 5\alpha \exp\left(-\frac{a(1-e^2)}{\lambda(1+e\cos\nu)}\right) \kappa\left(\frac{a(1-e^2)}{\lambda(1+e\cos\nu)}\right) \Phi_2\left(\frac{R_\Earth}{\lambda}\right) \right],
\end{multline}
where $\sigma(x) = 3+3x+2/3x^2-x^3/3$ and $\kappa(x)$ is defined above.

Eqs. (\ref{eq_rlg})-(\ref{eq_wlg}) clearly show the impact of a Yukawa interaction on a satellite's orbit. The first line of each equation provides the Newtonian part, while the Yukawa contribution is shown in the remaining terms. Not surprisingly, the Yukawa interaction impacts the perturbing force in a similar way it impacts the spherical harmonic coefficients (see Sect. \ref{ssect_y00y20}), through form factors and a complex radial dependence that couples an exponential decay with polynomials $\sigma(r/\lambda)$ and $\kappa(r/\lambda)$ which tend to maximise the effect for $r\sim \lambda$.

As aforementioned, it is well known that the $J_2$ zonal term sources a precession of the line of nodes through the ${\mathcal S}$ and ${\mathcal W}$ components of the perturbing force in the pure Newtonian case \cite{roy05}. Eqs. (\ref{eq_slg})-(\ref{eq_wlg}) show that a Yukawa interaction adds up to this effect. Its impact will depend on the strength $\alpha$ of the Yukawa interaction, but also on how its range $\lambda$ compares to the orbit's semi-major axis and to the radius of the Earth. We can therefore expect to measure different rates of precession for satellites orbiting at different altitudes.
Even for a homogeneous sphere ($f=0$), although the tangential components vanish ${\mathcal S} = {\mathcal W}=0$, the radial component remains affected by the form factor of the Earth: it simplifies to contain only the usual exponential decay coupled to the Earth's form factor.

We can therefore expect observable effects of the coupling of the Yukawa interaction to the shape of the Earth on the dynamics of satellites.
Hence, not taking the shape of the Earth into account to predict the very effects that are looked for to constrain a Yukawa interaction in orbit ends up in wrong predictions, and is likely to prevent reliable constraints.

In other words, it is incorrect to consider the perturbation due to the Yukawa interaction as a purely radial interaction sourced by a point-mass when working with satellite dynamics.
Most existing works that aim to constrain a Yukawa interaction with satellites dynamics focused on measuring the perigee precession under this incorrect assumption \cite{iorio02b,haranas11b, haranas11a, kolosnitsyn04, haranas16}. Nevertheless, although those works miss the contribution of the tangential components of the Yukawa interaction, we should note that since they focus their analyses on $\lambda\approx {\rm a \, few \,} R_\Earth$ (where $\Phi\approx 1$, see Fig. \ref{fig_Phi2}) their simplifying assumption only marginally affects the radial component of the perturbation.
However, if aiming to constrain short range Yukawa interaction, one has to take into account the fact that the Earth is an extended body, since in this regime the form factor is significantly greater than 1 and hence dramatically impacts the Gauss equations.

\section{Entanglement of geodesy and gravitation experiments in the Earth gravitational field} \label{sect_impact}

The discussion above shows that modified gravity affects the spherical harmonic coefficients of the Earth gravitational field, and in turn gravity observables (such as the motion of satellites).
Although this comes hardly as   a surprise, to the best of our knowledge, this  has never been seriously taken into account, neither to survey and invert the Earth gravity (to estimate the shape of the Earth) nor to constrain the Yukawa parameters in orbit.
In the former case, geophysicists assume that the Earth gravitational field is described by Newtonian gravity (hence, they ignore any Yukawa deviation altogether, see e.g. \cite{pail11, pavlis12}). In the latter case, for a Yukawa-like modification of gravitation, its effects on Keplerian parameters are most often computed under the assumption that the Yukawa acceleration is sourced by a point-mass Earth \cite{iorio02b,haranas11b,lucchesi11,kolosnitsyn04}, or at best by a uniform, spherical Earth \cite{berge18}.
And yet, a Yukawa deviation has explicit effects such as a dependence of spherical harmonic coefficients on the radial distance from the Earth.
Conversely, our imperfect knowledge of the Earth geometry may impact experimental constraints of the Yukawa parameters. 

Hagiwara \cite{hagiwara89} investigated the effect of a non-Newtonian contribution to the Earth gravitational field on geodesy experiments. He found that  non-Newtonian terms could safely be ignored to measure the Earth geoid. However, he was considering 1980's experiments precision, and as modern in-orbit gravity experiments such as GRACE, GOCE and GRACE-FO bring unprecedented precision on the measurement of the Earth gravitational field, it is timely to revisit his work. This is the purpose of this section and the next one. In this section, we first show the limitations that modified gravity brings to geodesy measurements, then those that geodesy uncertainties bring to tests of modified gravity, before giving recommendations on how to go beyond current limitations. Order-of-magnitude estimates are then given in Sect. \ref{sect_ofm}.

Fig. \ref{fig_flow} shows the entanglement between modified gravity (illustrated with a Yukawa interaction) and the shape of the Earth when testing gravity or measuring the Earth geometry with experiments in the Earth gravity. For simplicity, we still ignore relativistic effects, the influence of the Moon and other planets and the rotation of the Earth.
The system of interest is the Earth, whose geometry is coupled to a possible Yukawa potential; we aim to measure the Earth geometry and/or the Yukawa parameters. In the sense of Kant, they are noumenons (the ``true'' system), a priori not accessible to human senses, but which we can approximate by analyzing observable ``phenomenons''.
Those phenomenons can be as diverse as the value of the gravitational field acceleration ${\mathbf g}$, its gradient $[T]$, the equivalence principle, or the secular variations of Keplerian parameters (perigee drift $\Delta\omega$, regression of the line of nodes $\Delta\Omega$, variation of the eccentricity $\Delta e$).
Experiments provide us with ``measurements'' of those phenomenons, that are affected by statistical and systematic uncertainties. For instance, GOCE measured the gravitational gradient $[T]$, LAGEOS measured the perigee drift $\Delta\omega$ and the regression of the line of nodes $\Delta\Omega$, and MICROSCOPE tested the weak equivalence principle.
We may perform several measurements (each with its own expected value for the phenomenons under scrutiny, and each with its own uncertainties --stacked boxes in the figure), that we can then combine, e.g. simply by averaging their individual results ($\langle\dots\rangle$ is the ensemble average). 
Finally, those measurements can be used under some hypotheses and with some priors $\Pi$ on parameters to get estimates (possibly biased, and likely up to a given estimation error) of the underlying ``true'' parameters. The three boxes in the lower part of the figure show three different possible uses of Earth gravity measurements: geodesy (hypothesis {\bf H1} --sect. \ref{ssect_h1}), tests of gravity (hypothesis {\bf H2} --sect. \ref{ssect_h2}) and simultaneous geodesy and tests of gravity (hypothesis {\bf H3} --sect. \ref{ssect_h3}).

In this section, based on Fig. \ref{fig_flow}, we quantify the limitations on parameter estimations given some hypotheses. We do not try to be exhaustive and only give examples based on the measurement of the Newtonian spherical harmonic coefficients (Sect. \ref{ssect_h1}) and on the estimation of the Yukawa strength for a given range $\lambda$ from the combination of two satellite measurements (Sect. \ref{ssect_h2}). Sect. \ref{ssect_h3} discusses how to go beyond the limitations shown in Sect.  \ref{ssect_h1} and \ref{ssect_h2} for modified gravity experiments. Our discussion can be generalised to other observables (e.g. secular variations of Keplerian parameters), but we refrain from providing a full analysis of all possible experiments. Such analyses shall be presented in future works.

We note measurements and estimates with a hat: e.g., $\hat\alpha$ is the estimate of the Yukawa interaction strength. 
Modeled quantities are noted with a tilde: e.g., $\tilde{\alpha}$ is the strength of the Yukawa interaction for some a priori model.
We use the term ``prior'' loosely to denote an {\it a priori}, possibly subjective information on a parameter, and do not restrict its use to the Bayesian ``prior probability density function''.

\tikzstyle{decision} = [diamond, draw, fill=white,
    text width=4cm, text badly centered, node distance=3cm, inner sep=0pt]
\tikzstyle{block} = [rectangle, draw, fill=white, 
     text centered, rounded corners, minimum height=2em,
    execute at begin node={\begin{varwidth}{15em}},
   execute at end node={\end{varwidth}}]
\tikzstyle{squareblock} = [rectangle, draw, fill=white, 
    text width=7cm, text centered, minimum height=2em]
\tikzstyle{line} = [draw, -latex']
\tikzstyle{cloud} = [draw, ellipse,fill=white, text width=3.1cm, node distance=3cm, text badly centered,
    minimum height=2em]

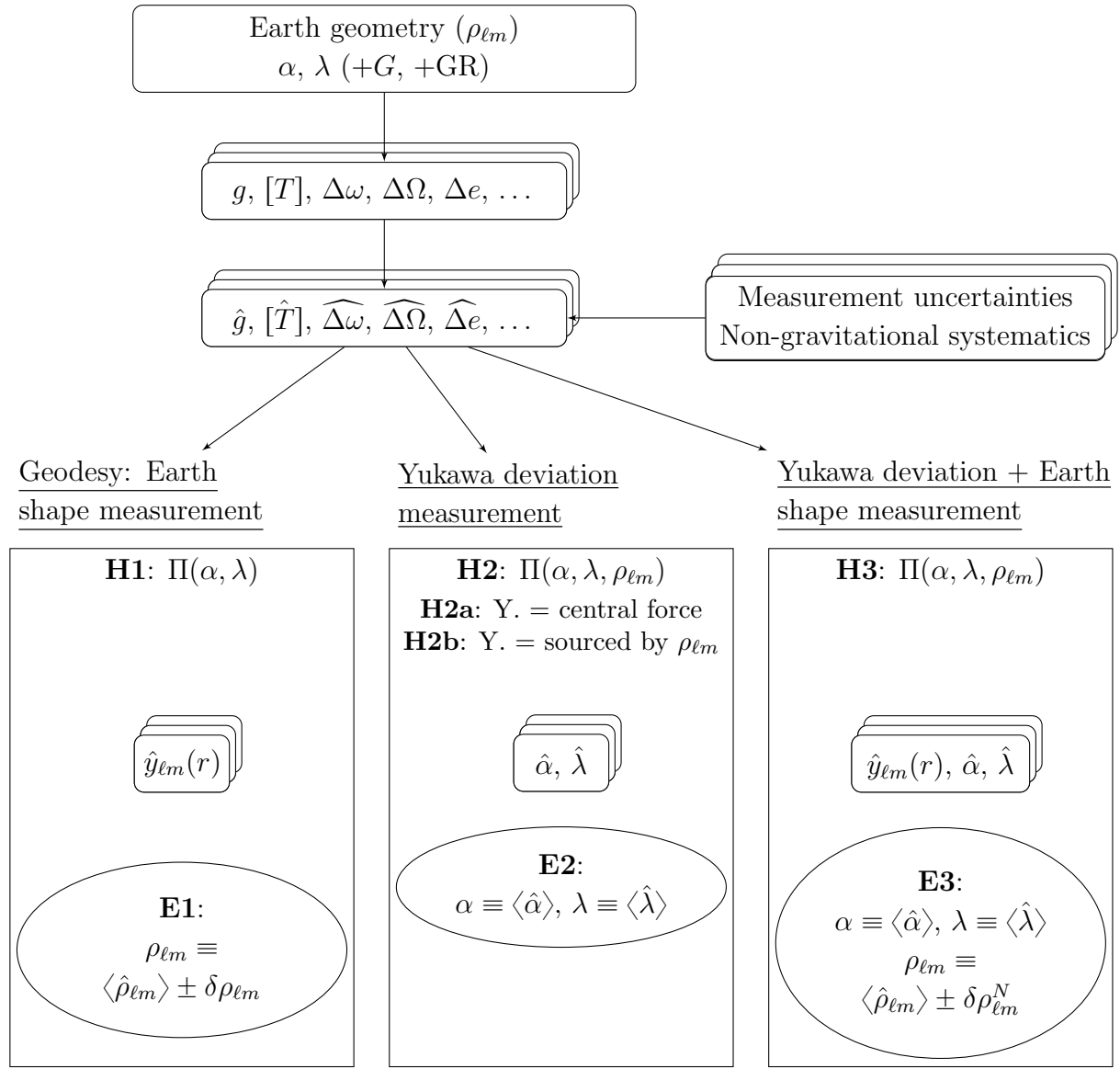
\begin{figure}
\begin{tikzpicture}
\node [rectangle, draw, fill=white, text centered, rounded corners, text width=7cm, minimum height=2em] at (10,0) (noumene) {Earth geometry ($\rho_{\ell m}$) \\ $\alpha$, $\lambda$ (+$G$, +GR)};

\node [rectangle, draw, fill=white, text centered, rounded corners, text width=5cm, minimum height=2em] (phenomenetmp) [below=1cm of noumene] {$\vec{g}$, $[T]$, $\Delta\omega$, $\Delta\Omega$, $\Delta e$, \dots};
\node [rectangle, draw, fill=white, text centered, rounded corners, text width=5cm, minimum height=2em] (phenomene2tmp) [above right=-0.7cm and -5.2cm of phenomenetmp] {};
\node [rectangle, draw, fill=white, text centered, rounded corners, text width=5cm, minimum height=2em] (phenomene3tmp) [above right=-0.7cm and -5.2cm of phenomene2tmp] {};
\node [rectangle, draw, fill=white, text centered, rounded corners, text width=5cm, minimum height=2em] (phenomenetmp) [above right=-0.7cm and -5.2cm of phenomenetmp] {};
\node [rectangle, draw, fill=white, text centered, rounded corners, text width=5cm, minimum height=2em] (phenomene) [below=1cm of noumene] {$g$, $[T]$, $\Delta\omega$, $\Delta\Omega$, $\Delta e$, \dots};

\node [rectangle, draw, fill=white, text centered, rounded corners, text width=5cm, minimum height=2em] (measurementtmp) [below=1cm of phenomene] {};
\node [rectangle, draw, fill=white, text centered, rounded corners, text width=5cm, minimum height=2em] (measurement2tmp) [above right=-0.7cm and -5.2cm of measurementtmp] {};
\node [rectangle, draw, fill=white, text centered, rounded corners, text width=5cm, minimum height=2em] (measurement3tmp) [above right=-0.7cm and -5.2cm of measurement2tmp] {};
\node [rectangle, draw, fill=white, text centered, rounded corners, text width=5cm, minimum height=2em] (measurementtmp) [above right=-0.7cm and -5.2cm of measurementtmp] {};
\node [rectangle, draw, fill=white, text centered, rounded corners, text width=5cm, minimum height=2em] (measurement) [below=1cm of phenomene] {$\hat{g}$, $\hat{[T]}$, $\widehat{\Delta\omega}$, $\widehat{\Delta\Omega}$, $\widehat{\Delta e}$, \dots};

\node [rectangle, draw, fill=white, text centered, rounded corners, text width=5.5cm, minimum height=3em] (systematicstmp) [right=2cm of measurement] {};
\node [rectangle, draw, fill=white, text centered, rounded corners, text width=5.5cm, minimum height=3em] (systematics2tmp) [above right=-1.1cm and -5.7cm of systematicstmp] {};
\node [rectangle, draw, fill=white, text centered, rounded corners, text width=5.5cm, minimum height=3em] (systematics3tmp) [above right=-1.1cm and -5.7cm of systematics2tmp] {};
\node [rectangle, draw, fill=white, text centered, rounded corners, text width=5.5cm, minimum height=3em] (systematicstmp) [above right=-1.1cm and -5.7cm of systematicstmp] {};
\node [rectangle, draw, fill=white, text centered, rounded corners, text width=5.5cm, minimum height=2em] (systematics) [right=2cm of measurement] {Measurement uncertainties \\ Non-gravitational systematics};

\node [rectangle, fill=white, text width=3.5cm, minimum height=2em] (geodesy) [below left=1.5cm and -1cm of measurement]  {\underline{Geodesy: Earth} \\ \underline{shape measurement}};
\node [rectangle, fill=white, text width=3.5cm, minimum height=2em] (yukawa) [right=1.7cm of geodesy] {\underline{Yukawa deviation} \\ \underline{measurement}};
\node [rectangle, fill=white, text width=3.5cm, minimum height=2em] (both) [right=1.7cm of yukawa] {\underline{Yukawa deviation + Earth} \\ \underline{shape measurement}};

\node [rectangle, draw, fill=white, text centered, text width=4.7cm, minimum height=7.5cm] (exp1) [below left=0.1cm and -5cm of geodesy] {};
\node (hyp1) [above=-0.7cm of exp1] {{\bf H1}: $\Pi(\alpha,\lambda)$};
\node [rectangle, draw, fill=white, text centered, rounded corners, text width=1.1cm, minimum height=2em] (est1tmp) [below=2cm of hyp1] {};
\node [rectangle, draw, fill=white, text centered, rounded corners, text width=1.1cm, minimum height=2em] (est12tmp) [above right=-0.7cm and -1.3cm of est1tmp] {};
\node [rectangle, draw, fill=white, text centered, rounded corners, text width=1.1cm, minimum height=2em] (est13tmp) [above right=-0.7cm and -1.3cm of est12tmp] {};
\node [rectangle, draw, fill=white, text centered, rounded corners, text width=1.1cm, minimum height=2em] (est1tmp) [above right=-0.7cm and -1.3cm of est1tmp] {};
\node [rectangle, draw, fill=white, text centered, rounded corners, text width=1.1cm, minimum height=2em] (est1) [below=2cm of hyp1] {$\hat{y}_{\ell m}(r)$};
\node [cloud] (est11) [below=1cm of est1] {{\bf E1}: \\ $\rho_{\ell m} \equiv \langle\hat{\rho}_{\ell m}\rangle \pm \delta\rho_{\ell m}$};

\node [rectangle, draw, fill=white, text centered, text width=4.7cm, minimum height=7.5cm] (exp2) [right=0.5cm of exp1] {};
\node (hyp2) [above=-0.7cm of exp2] {{\bf H2}: $\Pi(\alpha, \lambda, \rho_{\ell m})$};
\node (hyp2a) [below=-0.1cm of hyp2] {{\small{\bf H2a}: Y. = central force}};
\node (hyp2b) [below=-0.1cm of hyp2a] {{\small{\bf H2b}: Y. = sourced by $\rho_{\ell m}$}};
\node [rectangle, draw, fill=white, text centered, rounded corners, text width=1.1cm, minimum height=2em] (est2tmp) [below=1cm of hyp2b] {};
\node [rectangle, draw, fill=white, text centered, rounded corners, text width=1.1cm, minimum height=2em] (est22tmp) [above right=-0.7cm and -1.3cm of est2tmp] {};
\node [rectangle, draw, fill=white, text centered, rounded corners, text width=1.1cm, minimum height=2em] (est23tmp) [above right=-0.7cm and -1.3cm of est22tmp] {};
\node [rectangle, draw, fill=white, text centered, rounded corners, text width=1.1cm, minimum height=2em] (est2tmp) [above right=-0.7cm and -1.3cm of est2tmp] {};
\node [rectangle, draw, fill=white, text centered, rounded corners, text width=1.1cm, minimum height=2em] (est2) [below=1cm of hyp2b] {$\hat{\alpha}$, $\hat{\lambda}$};
\node [cloud] (est22) [below=0.5cm of est2] {{\bf E2}: \\ $\alpha \equiv \langle\hat{\alpha}\rangle$, $\lambda \equiv \langle\hat{\lambda}\rangle$};

\node [rectangle, draw, fill=white, text centered, text width=4.7cm, minimum height=7.5cm] (exp3) [right=0.5cm of exp2] {};
\node (hyp3) [above=-0.7cm of exp3] {{\bf H3}: $\Pi(\alpha, \lambda, \rho_{\ell m})$};
\node [rectangle, draw, fill=white, text centered, rounded corners, text width=2.3cm, minimum height=2em] (est3tmp) [below=2cm of hyp3] {};
\node [rectangle, draw, fill=white, text centered, rounded corners, text width=2.3cm, minimum height=2em] (est32tmp) [above right=-0.7cm and -2.5cm of est3tmp] {};
\node [rectangle, draw, fill=white, text centered, rounded corners, text width=2.3cm, minimum height=2em] (est33tmp) [above right=-0.7cm and -2.5cm of est32tmp] {};
\node [rectangle, draw, fill=white, text centered, rounded corners, text width=2.3cm, minimum height=2em] (est3tmp) [above right=-0.7cm and -2.5cm of est3tmp] {};
\node [rectangle, draw, fill=white, text centered, rounded corners, text width=2.3cm, minimum height=2em] (est3) [below=2cm of hyp3] {$\hat{y}_{\ell m}(r)$, $\hat{\alpha}$, $\hat{\lambda}$};
\node [cloud] (est32) [below=0.5cm of est3] {{\bf E3}: \\ $\alpha \equiv \langle\hat{\alpha}\rangle$, $\lambda \equiv \langle\hat{\lambda}\rangle$ \\ $\rho_{\ell m} \equiv \langle\hat{\rho}_{\ell m}\rangle \pm \delta\rho_{\ell m}^N$};

\path [line] (noumene) -- (phenomene);
\path [line] (phenomene) -- (measurement);
\path [line] (systematics) -- (measurement);
\path [line] (measurement) -- (geodesy);
\path [line] (measurement) -- (yukawa);
\path [line] (measurement) -- (both);
\end{tikzpicture}
\caption{Gravity experiments in the Earth gravitational field. The Earth (universal) geometry as described by the density spherical harmonics coefficients $\rho_{\ell m}$ (Eq. \ref{e.rho}), the gravitation constant and modified gravity parameters are unavailable to our senses (noumenons); they are the parameters of the theoretical model that can be used to try to know them. They can be observed through phenomenons (gravity acceleration, secular variations of satellites' osculating parameters) whose values depend on the values of the parameters of the model. Measurements provide us with estimates of those phenomenons (affected by statistical and systematic uncertainties). Depending on what hypotheses we make, we can use those measurements to estimate the parameters of the model: geodesy ({\bf H1}), modified gravity experiments ({\bf H2}), or both ({\bf H3}); the estimates {\bf E1}, {\bf E2}, {\bf E3} may be biased and known with some error depending on the hypothesis made.}
 \label{fig_flow}
\end{figure}

\subsection{Impact of Yukawa interaction on geodesy measurements ({\bf H1})} \label{ssect_h1}

Geodesy surveys are shown in Fig. \ref{fig_flow} by the left-hand-side panel. They aim to map the Earth geometry and mass distribution, as well as their time variations (estimates {\bf E1}) through the measurement of the static and/or variable gravitational field \cite{rummel11, pail11, tapley04, tapley05, reigber05, pail10, mayer06,pavlis12}: their goal is hence to estimate the spherical harmonics coefficients $\rho_{\ell m}(s)$ of the Earth density, as defined in Eq. (\ref{e.rho}).

Let us consider a satellite gravity survey performed at a distance $r$ from the centre of the Earth.
The survey provides estimates of the coefficients $y_{\ell m}$ of a spherical harmonic expansion, {\it a priori} independently of any underlying gravity model (as long as the Earth gravitational potentiel in a modified gravity model can be expanded in spherical harmonics): in our case, they contain both a Newtonian and a Yukawa contribution. 

A gravity model hypothesis {\bf H1} is then required to extract $\rho_{\ell m}$ from the measured $\hat{y}_{\ell m}$. It can either be a pure Newtonian field, or explicitly contain modified gravity. In the former case, $\hat{y}_{\ell m}$ is supposed to be given by Eq. (\ref{eq_ylmN}); in the latter case, it is supposed to be given by Eq. (\ref{e.ylmr}). 
If modified gravity is considered, the best way to proceed is the latter: invert Eq. (\ref{e.ylmr}) with some prior $\Pi(\alpha, \lambda)$ on the Yukawa interaction to obtain $\rho_{\ell m}$. However, to the best of our knowledge, all geodesy works use a Newtonian hypothesis and invert Eq. (\ref{eq_ylmN}) (e.g. \cite{grombein13, casenave16}). In this case, a non-zero Yukawa contribution will contaminate the analysis. A possible way to use the existing inversion codes based on Newtonian gravity is then to consider the Yukawa contribution as a systematic error, and just remove it from the estimated $\hat{y}_{\ell m}$ to then invert an estimated Newtonian coefficient. We thus assume a prior on the Yukawa parameters, which may be biased ($\E(\tilde\alpha) = \alpha + \delta\alpha$, $\E(\tilde\lambda) = \lambda + \delta\lambda$), where ($\alpha$, $\lambda$) are the true values and ($\delta\alpha$, $\delta\lambda$) are the prior's bias; we finally assume some uncertainty ($\Var(\tilde\alpha)$, $\Var(\tilde\lambda)$) on our prior. The Newtonian coefficient estimator then reads

\begin{equation}  \label{eq_ylm_estimator}
\hat{y}_{\ell m}^N = \hat{y}_{\ell m} - \tilde{\alpha}\widetilde{f_\ell(r,\lambda) q_{\ell m}^Y/q_{00}^N}
\end{equation}
where the function $f_{\ell}(r,\lambda)={\mathcal B}_l(r/\lambda)/(2\ell+1)$ encapsulates the prior on $\lambda$ (which affects the gravitational field model in a non-trivial way that we do not attempt to compute) and the quantities
\begin{eqnarray}
q_{\ell m}^N &=& \int s^2 \left(\frac{s}{R_\Earth}\right)^\ell \rho_{\ell m}(s) {\rm d}s \\
\label{eq_qy2}
q_{\ell m}^Y &=& \int s^2 \left(\frac{s}{R_\Earth}\right)^\ell \rho_{\ell m}(s) {\mathcal A}_l\left(\frac{s}{\lambda}\right) {\rm d}s
\end{eqnarray}
are the integrals over the volume of the Earth that appear in Eqs. (\ref{eq_ylmN})-(\ref{eq_ylmY}), whose dependence on the geoid and mass density are not yet important, but will be developed below. In the remainder of this section, we do not attempt to obtain $\rho_{\ell m}$, but use $\hat{y}_{\ell m}^N$ as a proxy. Note that trivially, $q_{00}^N=M_\Earth$.

The expected value and variance of the estimator (\ref{eq_ylm_estimator}) are
\begin{equation} \label{eq_ylm_ev}
\E(\hat{y}_{\ell m}^N) = y_{\ell m}^N - \alpha \delta\left[ f_{\ell}(r,\lambda) \frac{q_{\ell m}^Y}{q_{00}^N} \right] - \delta\alpha \left( f_{\ell}(r, \lambda) \frac{q_{\ell m}^Y}{q_{00}^N}\right) - \delta\alpha \delta\left[ f_{\ell}(r, \lambda) \frac{q_{\ell m}^Y}{q_{00}^N} \right]
\end{equation}
and
\begin{equation} \label{eq_ylm_var}
\Var(\hat{y}_{\ell m}^N) = \Var(\hat{y}_{\ell m}) + \tilde\alpha^2 \Var(\widetilde{ f_{\ell}(r, \lambda) \frac{q_{\ell m}^Y}{q_{00}^N}}) + \left(  \widetilde{f_{\ell}(r, \lambda) \frac{q_{\ell m}^Y}{q_{00}^N}}\right)^2 \Var(\tilde\alpha) + \Var(\tilde\alpha) \Var\left(\widetilde{f_{\ell}(r, \lambda) \frac{q_{\ell m}^Y}{q_{00}^N}}\right)
\end{equation}
where we assume that the measurement itself is unbiased ($\E(\hat{y}_{\ell m}) = y_{\ell m}$), that it is independent of the prior on the Yukawa interaction and where, for clarity, we ignore all other possible systematic errors (e.g., solar radiation pressure, atmospheric drag, tidal effects, mass motion on the Earth surface...).

We should also note that beside a prior on the Yukawa interaction, a prior on the Earth mass distribution is implicitly used in $\widetilde{f_\ell(r,\lambda) q_{\ell m}^Y/q_{00}^N}$ (see Eq. \ref{eq_qy2}). It may as well come from experiments completely independent of the gravitational field (e.g. seismology surveys) or from gravitational field measurements, through the estimation of the spherical harmonic coefficients. In the latter case, the problem becomes non-linear, since the prior is based on knowledge similar to what we wish to measure. Although we should keep that in mind, we ignore this aspect and assume that the prior is indeed uncorrelated with the measurement.

Eqs. (\ref{eq_ylm_ev})-(\ref{eq_ylm_var}) allow us to conclude on the effect of a Yukawa interaction on the estimation of $\hat{y}_{\ell m}^N$. Eq. (\ref{eq_ylm_ev}) shows that a prior on the Yukawa interaction too far from the real characteristics of the Yukawa interaction (or simply ignoring the possibility of a Yukawa interaction if it actually exists) leads to a biased estimate of the Newtonian contribution to the Earth gravitational field.  Eq. (\ref{eq_ylm_var}) shows that a physically-motivated prior increases the variance of the estimator (i.e., which is not anymore equal to the variance of the measured $\hat{y}_{\ell m}$ as when ignoring the possibility of a Yukawa interaction): this is the price to pay to have an unbiased estimate $\hat{y}_{\ell m}^N$.
With those observations in mind, one must be aware that using (as usual) the Newtonian framework for geodesy (i.e. assuming $\alpha=0$ and $\delta\alpha=0$) may lead to biased estimations of the Earth geometry if in reality $\alpha \neq 0$; furthermore, in this case, the uncertainties on the estimates are underestimated.

We can also note that the bias and variance of the $\hat{y}_{\ell m}^N$ estimator depend on the distance of the experiment to the centre of the Earth through the radial dependence of $f_\ell(r,\lambda)$. Therefore, if modified gravity is real, then under the incorrect hypothesis that gravity is purely Newtonian (in which case it is assumed that the measured coefficients $\hat{y}_{\ell m}=\hat{y}_{\ell m}^N$), we may expect that different estimators $\hat{y}_{\ell m}^N$ obtained at different altitudes will be inconsistent, each with a non-zero bias and an underestimated variance coming from an incorrect hypothesis, even if the measurements are perfect.
This is reminiscent of the inconsistent measurements of the $y_{20}$ parameter between the GOCE-only, GRACE-only and EIGEN-6C models mentioned in the introduction. Answering the question of whether the tension between those measurements stems from data analyses or from the presence of a Yukawa interaction is beyond the scope of this paper, but could be done by re-analyzing all the concerned data with a model that takes into account the possible presence of a Yukawa potential and using realistic priors on the Yukawa interaction. \footnote{Quick-and-dirty constraints of the Yukawa interaction from the 700 $\sigma$ tension mentioned in the introduction provide results highly inconsistent with published constraints. The most likely reason is an incorrect error analysis from gravity surveys. See Sect. \ref{ssect_h2} for a discussion on how to constrain the Yukawa interaction by combining GOCE and GRACE measurements.}

Moreover, as shown by Eq. (\ref{eq_ylm_var}), the Yukawa interaction increases the variance of the $\hat{y}_{\ell m}$ estimator for non-circular orbits through the $f_\ell(r,\lambda)$'s dependence on $r$. This increase is also non-zero when combining several measurements made with satellites at different altitudes.

We give order-of-magnitude estimates of the effect of a non-zero Yukawa interaction on the $y_{\ell m}$ coefficients in Sect. \ref{sect_ofm}.

\subsection{Impact of Earth's gravity and shape errors on the measurement of Yukawa parameters ({\bf H2})} \label{ssect_h2}

Tests of gravity are shown by the middle panel of Fig. \ref{fig_flow}. The aim is to measure the Yukawa interaction parameters {\bf E2} (other applications can be e.g. to measure any relativistic effect) under some hypotheses {\bf H2} and priors on the Yukawa parameters ($\alpha$, $\lambda$) and/or the Earth shape ($\rho_{\ell m}$) and/or a direct measurement of the gravitational field with no explicit discrimination between the Newtonian and Yukawa contributions ($y_{\ell m}$). 
An extra prior consists in how the Yukawa interaction is modeled. Independently of the assumptions on the Newtonian gravitational field, we can either assume that it is sourced by a point-mass-like Earth ({\bf H2a}) or by the full, complex shape of the Earth ({\bf H2b}). As aforementioned, to our knowledge, most works \cite{iorio02b, haranas11b, lucchesi11, haranas11a} use the {\bf H2a} hypothesis, when a handful either briefly discuss or effectively use a spherical Earth (simplified {\bf H2b} hypothesis --\cite{berge18, haranas16}), but we could not find any use of a non-spherical Earth to constrain a Yukawa interaction.
Similarly, to our knowledge, no prior on $\alpha$ nor $\lambda$ has ever been used, although it is common practice to consider at least the measured $y_{20}$ zonal term of the Earth gravitational field to correct for its Newtonian contribution.

Several observables can be used to constrain a Yukawa interaction with experiments in the Earth gravitational field. Published works use the secular variation of Keplerian parameters of orbiting satellites like LAGEOS I \& II \cite{iorio02b, haranas11b, lucchesi11, haranas11a} under the {\bf H2a} hypothesis (the Yukawa interaction is sourced by a point-mass Earth), or the measured (absence of) violation of the equivalence principle \cite{berge18}.
Given the link between the spherical harmonics coefficients and the Yukawa interaction, we could also think of constraining the Yukawa parameters directly from the measured $\hat{y}_{\ell m}$, either from a single experiment or from a combination of experiments and/or different $\hat{y}_{\ell m}$. To the best of our knowledge, such an analysis, based on hypothesis {\bf H2b}, has never been performed.
As already mentioned, we do not try to be exhaustive, and will only provide details for one possible way to constrain the Yukawa parameter. Therefore, in the remainder of this section, we propose to combine the $\hat{y}_{\ell m}$ coefficient measured by two experiments at different altitudes (for a given pair ($\ell$, $m$)) and show how it can shed light on the Yukawa interaction.

Let us assume that $y_{\ell m}$ is estimated by two different experiments at distances $r_1$ and $r_2$ from the centre of the Earth, to provide two estimators $\hat{y}_{\ell m,1}$ and $\hat{y}_{\ell m,2}$.
Using Eq. (\ref{e.ylmr}), we can form the following estimator of $\alpha$, for a given range $\lambda$, from the difference between the two $\hat{y}_{\ell m}$ estimators: 
\begin{equation} \label{eq_alpha_estimator}
\hat{\alpha}_{\ell m} = \frac{q_{00}^N(\tilderho, \tildeh)}{[f_\ell(r_1,\lambda) - f_\ell(r_2,\lambda)] q_{\ell m}^Y(\tilderho, \tildeh)} (\hat{y}_{\ell m,1} - \hat{y}_{\ell m,2}),
\end{equation}
where the functions $q_{\ell m}^N$ and $q_{\ell m}^Y$ were defined above and $q_{00}^N=M_\Earth$; we now write their explicit dependence on the mass density distribution $\rho({\mathbf x})$ and on the geoid $h({\mathbf x})$ --just another way to see the information contained in $\rho_{\ell m}(s)$. This estimator is clearly Earth-model-dependent.
Although different in its purpose, it is related to Wagner \& McAdoo's error factor \cite{wagnermcadoo12}; in that case, it serves as a way to calibrate different (Newtonian) gravitational field models, while we treat it as a measure of non-Newtonian deviations. We should also note that a better estimator would be to average (\ref{eq_alpha_estimator}) over all ($\ell$,$m$) pairs, but for the sake of clarity, we only discuss (\ref{eq_alpha_estimator}) in the following.

The prior on the Earth model propagates in a non-trivial way to a bias and uncertainty on the $q_{\ell m}$ functions. We do not try to perform this computation (which should be done numerically and requires specifying a model for the Earth), but assume that instead of dealing with priors on the mass distribution and the geoid, we have (biased) priors on the $q_{\ell m}$ functions (for clarity, we drop the $\rho$ and $h$ dependences), such as $\E(\tilde{q}_{\ell m}) = q_{\ell m} + \delta q_{\ell m}$, which applies both to the Newtonian and to the Yukawa contributions to the gravitational field. Additionally, we assume that the Earth model is based only on data independent of the gravitational field (e.g. seismology surveys); otherwise, the problem is non-linear since (as seen in Sect. \ref{ssect_h1}) the model depends on our knowledge of the Yukawa interaction.

Under those hypotheses, the expected value of the $\hat\alpha_{\ell m}$ estimator is
\begin{multline} \label{eq_expvalalpha_ylm_diff}
\E(\hat\alpha_{\ell m}) = \frac{q_{\ell m}^Y}{q_{\ell m}^Y + \delta q_{\ell m}^Y} \left\{ \alpha_{\ell m} \right. \\
+ \left. \frac{q_{00}^N}{[f_{\ell}(r_1,\lambda) - f_{\ell}(r_2,\lambda) ]q_{\ell m}^Y} \left[ \left(1 + \frac{\delta q_{00}^N}{q_{00}^N}\right) (\delta y_{\ell m,1} - \delta y_{\ell m,2})
+ \frac{\delta q_{00}^N}{q_{00}^N} (y_{\ell m,1} - y_{\ell m,2}) \right] \right\}
\end{multline}
We should note that the bias in the measured $\hat{y}_{\ell m}$ may not be the same for the two satellites.
It is then apparent that a biased $\hat{y}_{\ell m}$ contributes an additive bias to $\hat\alpha_{\ell m}$, while a biased Earth model contributes both an additive and a multiplicative bias to $\hat\alpha_{\ell m}$.
These biases can be minimised by minimizing $\delta y_{\ell m}$ and $\delta q_{\ell m}^{N,Y}$ (i.e. improving the accuracy of the $y_{\ell m}$ measurement and of the Earth model).

The variance of this estimator can then be shown to be
\begin{multline} \label{eq_varalpha_ylm_diff}
\Var(\hat{\alpha}_{\ell m}) = \left(\frac{q_{00}^N + \delta q_{00}^N}{[f_{\ell}(r_1,\lambda)  - f_{\ell}(r_2,\lambda)] (q_{\ell m}^Y + \delta q_{\ell m}^Y)}\right)^2 \\
\times \left[\left(1 + \frac{\Var(\tilde{q}_{00}^N)}{\left(q_{00}^N + \delta q_{00}^N\right)^2} -2 \frac{\Cov(\tilde{q}_{00}^N, \tilde{q}_{\ell m}^Y)}{(q_{00}^N + \delta q_{00}^N)(q_{\ell m}^Y + \delta q_{\ell m}^Y)} + \frac{\Var(\tilde{q}_{\ell m}^Y)}{\left(q_{\ell m}^Y + \delta q_{\ell m}^N\right)^2}
\right) \left[\Var(\hat{y}_{\ell m,1}) + \Var(\hat{y}_{\ell m,2})\right] \right. \\
\left. +  \left(\frac{\Var(\tilde{q}_{00}^N)}{\left(q_{00}^N + \delta q_{00}^N\right)^2} -2 \frac{\Cov(\tilde{q}_{00}^N, \tilde{q}_{\ell m}^Y)}{(q_{00}^N + \delta q_{00}^N)(q_{\ell m}^Y + \delta q_{\ell m}^Y)} + \frac{\Var(\tilde{q}_{\ell m}^Y)}{\left(q_{\ell m}^Y + \delta q_{\ell m}^N\right)^2} \right) \left(y_{\ell m,1} + \delta y_{\ell m,1} - y_{\ell m,2} - \delta y_{\ell m,2}\right)^2 \right].
\end{multline}

Now assuming that the spherical harmonics coefficients measurements are unbiased and the biases on the Earth model are small, then at first order, Eq. (\ref{eq_varalpha_ylm_diff}) reads
\begin{multline} \label{eq_varalpha_ylm_diff1}
\Var(\hat{\alpha}_{\ell m}) \approx \left(\frac{q_{00}^N}{[f_{\ell}(r_1,\lambda)  - f_{\ell}(r_2,\lambda)] q_{\ell m}^Y}\right)^2 \left(1 - 2 \frac{\delta q_{\ell m}^Y}{q_{\ell m}^Y} + 2 \frac{\delta q_{00}^N}{q_{00}^N}\right)  \\
\times \left\{ \left[ 1 + \frac{\Var(\tilde{q}_{00}^N)}{\left(q_{00}^N\right)^2} \left(1 - 2 \frac{\delta q_{00}^N}{q_{00}^N}\right) -2 \frac{\Cov(\tilde{q}_{00}^N, \tilde{q}_{\ell m}^Y)}{q_{00}^Nq_{\ell m}^Y} \left(1 - \frac{\delta q_{\ell m}^Y}{q_{\ell m}^Y} - \frac{\delta q_{00}^N}{q_{00}^N}\right) \right. \right.\\
\left. + \frac{\Var(\tilde{q}_{\ell m}^Y)}{\left(q_{\ell m}^Y\right)^2}\left(1 - 2 \frac{\delta q_{\ell m}^Y}{q_{\ell m}^Y}\right) \right] \left[\Var(\hat{y}_{\ell m,1}) + \Var(\hat{y}_{\ell m,2})\right] \\
+ \left[\frac{\Var(\tilde{q}_{00}^N)}{\left(q_{00}^N\right)^2} \left(1 - 2 \frac{\delta q_{00}^N}{q_{00}^N}\right) -2 \frac{\Cov(\tilde{q}_{00}^N, \tilde{q}_{\ell m}^Y)}{q_{00}^Nq_{\ell m}^Y} \left(1 - \frac{\delta q_{\ell m}^Y}{q_{\ell m}^Y} - \frac{\delta q_{00}^N}{q_{00}^N}\right) \right.\\
\left. \left. + \frac{\Var(\tilde{q}_{\ell m}^Y)}{\left(q_{\ell m}^Y\right)^2}\left(1 - 2 \frac{\delta q_{\ell m}^Y}{q_{\ell m}^Y}\right) \right] \left(y_{\ell m,1} - y_{\ell m,2}\right)^2 \right\}
\end{multline}

Similarly to what happened for the expected value, Eqs. (\ref{eq_varalpha_ylm_diff})-(\ref{eq_varalpha_ylm_diff1}) show that several contributions make up the variance of $\hat{\alpha}_{\ell m}$: the variance and bias of the measured $\hat{y}_{\ell m}$ as well as the uncertainties and biases on the Earth model used for the analysis (which go in the {\bf H2b} hypotheses of Fig. \ref{fig_flow}). In particular, a biased Earth model affects the variance of $\hat{\alpha}_{\ell m}$ in a non-trivial way, whereby the bias on the mass (remember that $q_{00}^N=M_\Earth$) may or may not be counterbalanced by the bias on $q_{\ell m}^Y$, so that the impact of the Earth model bias will depend on the multipole ($\ell$, $m$) considered for the analysis. However, exploring the details of this question is far beyond the scope of this paper.

Eq. (\ref{eq_expvalalpha_ylm_diff})-(\ref{eq_varalpha_ylm_diff1}) are the bases for a signal-to-noise analysis to optimise the significance of the estimation of $\alpha$ for a given $\lambda$, for a given mission made of two satellites; for example, given a model of the Earth, it allows us to define the satellites' altitude or the optimal $y_{\ell m}$ that should be used to constrain $\alpha$ with a given precision and accuracy. Furthermore, by comparing both contributions to the variance, it directly provides clues about the limitations brought by our imperfect knowledge of the Earth, and can therefore set a lower bound on the measurement precision and accuracy required to reach a given precision on the Yukawa parameters.

Such an analysis, linked to a given mission concept, should be done numerically, and goes beyond the scope of this paper.
Nevertheless, we can give some crude order of magnitude estimate.
For instance, ignoring the covariance between $q_{00}^N$ and $q_{\ell m}^Y$ and the bias on the Earth model, we can compare the relative contribution to the variance of the $\hat{y}_{\ell m}$ measurements and of our imperfect model of the Earth. For instance, considering $(\ell,m)=(2,0)$, and assuming that $\Var(\tilde{q}_{20}^Y)/q_{20}^Y \approx \Var(y_{20})/y_{20} \approx 10^{-16}$ \cite{pail11} and that $\Var(q_{00}^N)/q_{00}^N = \Var(M_\Earth)/M_\Earth \approx 10^{-8}$, we find that the variance of the Yukawa strength estimator is limited by the $y_{20}$ measurement if $\Var(\hat{y}_{20}) > 10^{-8} (y_{20,1} - y_{20,2})$.
Further assuming that $y_{20,1} - y_{20,2} \sim 10^{-13}$\footnote{For illustrative purpose. Given the current experimental limits on the Yukawa interaction, this value is about the maximum that could still be measured by two satellites, at altitudes of 250 km and 2500 km, for $\lambda \sim 1.2\times10^5$m --see Sect. \ref{sect_ofm}}, we find that unless we have an improved Earth model, the error on $\hat\alpha$ will saturate as soon as we measure $y_{20}$ with a precision (square root of the variance) better than $10^{-17}$. As the current measured uncertainty on $y_{20}$ is of order $10^{-12}$ \cite{mayer06}, this crude order of magnitude estimate shows that were we to use the difference of $J_2$ between two satellite measurements made at different altitudes, we can improve the instrumental precision by five orders of magnitude before our constrain on $\alpha$ would become dominated by the Earth model. The limitation due to the Earth model would be even farther down if $y_{20,1} - y_{20,2}$ happens to be less than our assumed $10^{-13}$.

As already mentioned, we consider the 700 $\sigma$ tension between the $y_{20}$ coefficient measured by GOCE and GRACE dubious, and hence refrain from deriving any constrain on the Yukawa interaction, since the most likely cause for the tension is linked to error analyses.
We present a better motivated example for the $y_{20}$ case, in a homogeneous Earth model, in Sect. \ref{sect_ofm}.

Although similar considerations could be made when constraining a Yukawa interaction from the measurement of satellite orbits and secular variations of Keplerian parameters, we only mention that given the dependence of the Lagrange-Gauss equations on the shape of the Earth, constraints will undoubtely be impacted by the model of the Earth used for the analysis.

\tikzstyle{decision} = [diamond, draw, fill=white,
    text width=1.8cm, text badly centered, node distance=3cm, inner sep=0pt]
\tikzstyle{block} = [rectangle, draw, 
     text centered, rounded corners, minimum height=2em,
    execute at begin node={\begin{varwidth}{15em}},
   execute at end node={\end{varwidth}}]
\tikzstyle{squareblock} = [rectangle, draw, fill=white, 
    text width=7cm, text centered, minimum height=2em]
\tikzstyle{line} = [draw, -latex']
\tikzstyle{dashedline} = [draw, dashed, -latex']
\tikzstyle{cloud} = [draw, ellipse,node distance=3cm, minimum height=2em]

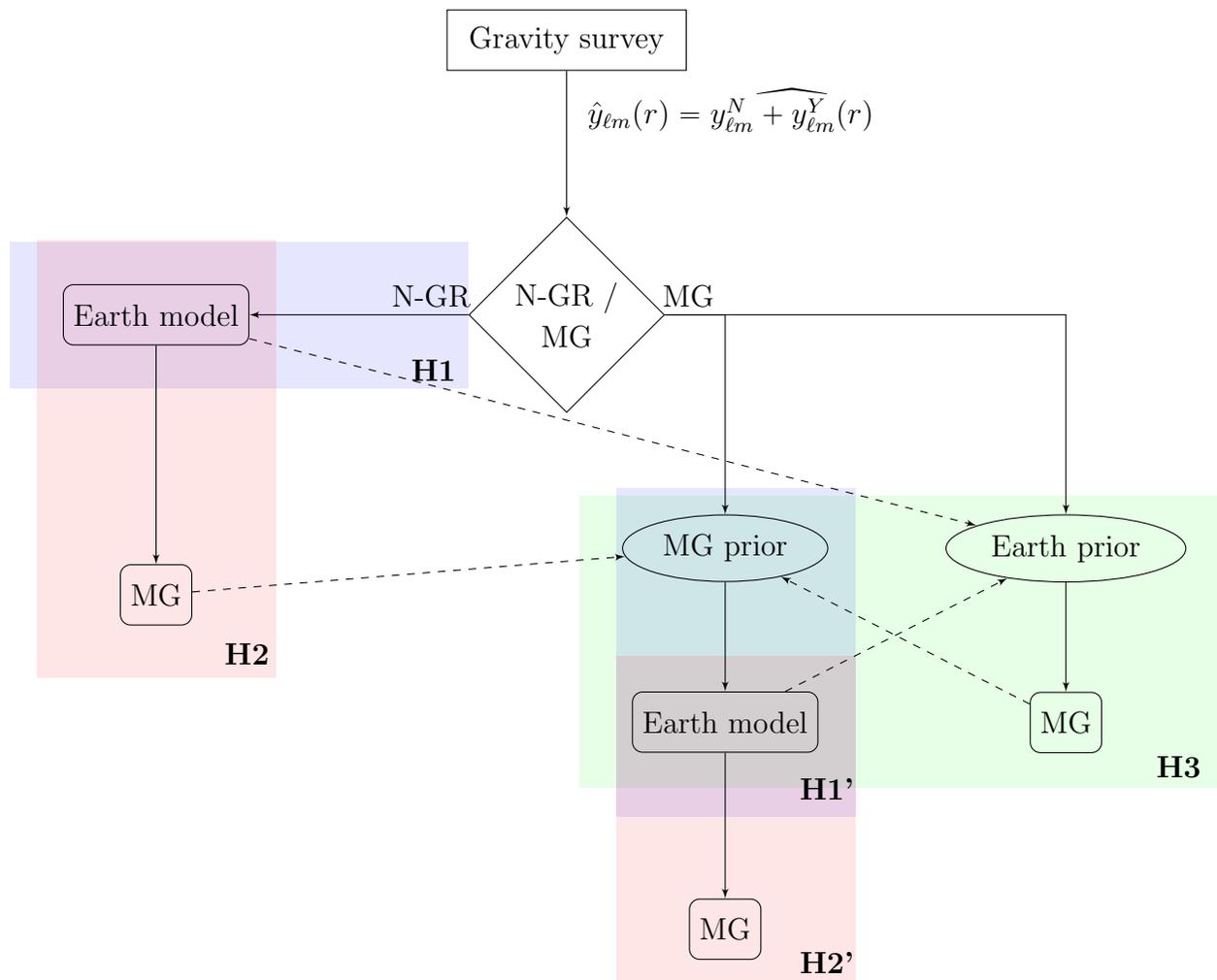
\begin{figure}
\begin{center}
\begin{tikzpicture}
\node [rectangle, draw, fill=white, text centered, text width=3cm, minimum height=2em] (survey) {Gravity survey};
\node [decision, below=2cm of survey] (frame) {N-GR / MG};
\node [rectangle, dotted, dashed, fill opacity=0.1, fill=green, text width=8.5cm, minimum height=4cm, below right = 1.8cm and -0.5cm of frame] (h3) {};
\node [rectangle, dotted, dashed, fill opacity=0.1, fill=blue, text width=6cm, minimum height=2cm, left = 0cm of frame] (h1) {};
\node [rectangle, dotted, dashed, fill opacity=0.1, fill=red, text width=3cm, minimum height=6cm, below left =-1.7cm and 3.3cm of frame] (h2) {};
\node [rectangle, dotted, dashed, fill opacity=0.1, fill=blue, text width=3cm, minimum height=4.5cm, below right =1.7cm and 0cm of frame] (h1beyond) {};
\node [rectangle, dotted, dashed, fill opacity=0.1, fill=red, text width=3cm, minimum height=4.5cm, below right =4cm and 0cm of frame] (h2beyond) {};
\node [block, left=3cm of frame] (emodel_h1) {Earth model};
\node [block, below=3cm of emodel_h1] (mg_h2present) {MG};
\node [cloud, below right=2.2cm and 5cm of frame] (eprior_h3) {Earth prior};
\node [cloud, below right=2.2cm and 0.5cm of frame] (mgprior_h3) {MG prior};
\node [block, below=1.5cm of eprior_h3] (mg_h3) {MG};
\node [block, below=1.5cm of mgprior_h3] (emodel_h3) {Earth model};
\node [block, below=2cm of emodel_h3] (mg_h2beyond) {MG};

\node [rectangle, below right= 0.1cm and -1.5cm of survey] (ylm) {$\hat{y}_{\ell m}(r) = \widehat{y_{\ell m}^N + y_{\ell m}^Y(r)}$};
\node [rectangle, above right= -0.7cm and 0.5cm of frame] (MG_choice) {MG};
\node [rectangle, above left= -0.7cm and 0.5cm of frame] (N_choice) {N-GR};
\node [rectangle, below right=-0.1cm and 0.6cm of mg_h3] (h3text) {{\bf H3}};
\node [rectangle, below left=-0.2cm and 0.7cm of frame] (h1text) {{\bf H1}};
\node [rectangle, below right=0.1cm and 0.3cm of mg_h2present] (h2text) {{\bf H2}};
\node [rectangle, below right=0.2cm and -0.4cm of emodel_h3] (h1beyondtext) {{\bf H1'}};
\node [rectangle, below=1.8cm of h1beyondtext] (h1beyondtext) {{\bf H2'}};

\path [line] (survey) -- (frame);
\path [line] (frame) -- (emodel_h1);
\path [line] (frame) -| (eprior_h3);
\path [line] (frame) -| (mgprior_h3);
\path [line] (eprior_h3) -- (mg_h3);
\path [line] (mgprior_h3) -- (emodel_h3);
\path [line] (emodel_h3) -- (mg_h2beyond);
\path [line] (emodel_h1) -- (mg_h2present);
\path [dashedline] (emodel_h3) -- (eprior_h3);
\path [dashedline] (mg_h3) -- (mgprior_h3);
\path [dashedline] (mg_h2present) -- (mgprior_h3);
\path [dashedline] (emodel_h1) -- (eprior_h3);

\end{tikzpicture}
\end{center}
\caption{Flowchart for modified gravity experiments in the Earth gravitational field with at least one aspect of data analysis based on an external gravity model. An external gravity survey provides the measured coefficients of a spherical harmonic expansion; they contain the contributions from the Newtonian (Earth shape) and the modified gravitational fields. A model is then decided upon to extract information from those $\hat{y}_{\ell m}$ coefficients: either we assume gravity is Newtonian (N-GR), or we include modified gravity in the model (MG). In the former case, we can derive a (possibly biased, if modified gravity actually exists) model of the Earth (Fig. \ref{fig_flow}'s {\bf H1} hypothesis), from which we can constrain (possibly biased) estimators for modified gravity ({\bf H2} hypothesis). In the latter case, priors on modified gravity allow for an (unbiased) model of the Earth ({\bf H1'} hypothesis), from which (unbiased) constraints on modified gravity can be drawn ({\bf H2'} hypothesis). Using both priors on modified gravity and on the shape of the Earth, we can get an updated Earth model and modified gravity constraints simultaneously ({\bf H3} hypothesis). Dashed lines show the interplay between priors and measurements, and show that we can iterate to improve upon the analysis; a Bayesian approach is even better in the {\bf H3} case.}
 \label{fig_flow2}
\end{figure}

\subsection{Going beyond current hypotheses and analyses in modified gravity experiments} \label{ssect_h3}

The discussion above allowed us to identify limitations inherent to current experiments in geodesy and modified gravity in the vicinity of the Earth. For instance, although Earth gravity surveys are almost model-independent (apart from the facts that it is assumed that the gravitational field can be expanded on a spherical harmonics basis and that by definition $y_{00}=1$ and is not estimated), geodesy experiments must choose a model to invert a gravity map into a model of the Earth. On the one hand, ignoring the possibility for modified gravity may end up on a biased Earth model. 
On the other hand, modified gravity experiments based on an explicit Earth model (like the estimator presented in Sect. \ref{ssect_h2}) are impacted by a biased and imprecise model of the Earth. This is most likely the case if they rely on a model derived from a global gravity survey (small scale ground tests relying on the modeling of the laboratory surroundings are less prone to this kind of errors).
This process is shown by the left arm of the flowchart depicted in Fig. \ref{fig_flow2}: from a gravity survey and its measured $y_{\ell m}$ coefficients, a (biased) model of the Earth is derived under the assumption that gravity has only a Newtonian contribution ({\bf H1} hypothesis of Fig. \ref{fig_flow} with no prior on modified gravity), then (biased) constraints on modified gravity are derived ({\bf H2} hypothesis).

As the bias and uncertainty of the Earth model propagate to the constraints on modified gravity only if those constraints are model-dependent, a possibility to avoid this limitation is to define model-independent constraints. Other combinations of spherical harmonic coefficients may be thought of that, in principle, cancel the contributions from the model of the Earth. For instance, the ratio of $\hat{y}_{20}$ measured by two satellites is independent of the Earth details. However, this is true only for spherical harmonics coefficients as defined in Sect. \ref{sect_general}, where $y_{00}$ is not universally equal to 1 but depends on the distance to the Earth. On the contrary, spherical harmonic coefficients provided by gravity surveys give $y_{00}=1$ by definition. This discrepancy, beside implicitly combining inconsistent models, would force us to renormalise our $y_{20}$ by $y_{00}$, making them effectively depend on the Earth characteristics, with a different dependence for both satellites. Therefore, given the current gravity surveys measurements, it is not possible to avoid Earth model uncertainties.

Those difficulties arise in the {\bf H2b} hypothesis, whereby we compute the Yukawa contribution sourced by the shape of the Earth. Most published constraints on the Yukawa interaction use the {\bf H2a} hypothesis and ignore the shape of the Earth altogether, besides the effect of the Earth flattening. This is no better than using a biased Earth model, since it amounts to using inconsistent gravity models (extended Earth for the Newtonian part of the gravity field, and point-mass Earth for the Yukawa contribution). We then claim that the {\bf H2a} hypothesis should be dropped and replaced by the {\bf H2b} hypothesis.

At this point, it should be clear that we are currently facing two main problems. The first one is the use of inconsistent models in geodesy and in modified gravity experiments. The second one is the entanglement of geodesy and modified gravity experiments, which ends up in non-linear error propagation and interdependent models, priors and constraints.

A natural solution to the inconsistent models problem is simply to derive geodesy results from gravity surveys with modified gravity in mind. Instead of considering the measured spherical harmonics coefficients as pure representations of the (Newtonian) geometry, the contribution from modified gravity should be taken into account. This is shown by the {\bf H1'} frame in the right arm of Fig. \ref{fig_flow2}'s flowchart. By assuming a gravity model to which both the Newtonian and the Yukawa interaction contribute and using an appropriate prior on the Yukawa parameters, the Earth model becomes unbiased, though its variance is increased, as shown in Sect. \ref{ssect_h1}. Then, we can safely use this Earth model to derive unbiased constraints on modified gravity ({\bf H2'} frame). The dashed line between {\bf H2} and the {\bf H1'} prior on modified gravity show how existing constraints on the Yukawa interaction can readily be used and marginalised over to obtain a better Earth model, from which updated constraints on the Yukawa interaction can be derived.

Another possibility is not only to derive geodesy results with modified gravity in mind, but to perform geodesy and modified gravity experiments simultaneously. This is shown by the {\bf H3} hypothesis in Figs. \ref{fig_flow} and \ref{fig_flow2}. This option has the advantage to allow for the use of the same data set for both analyses, thereby lowering the risk of errors coming from incompatible data sets. Moreover, as shown by the dashed lines in the {\bf H3} frame of Fig. \ref{fig_flow2}, such a solution allows for easy iterations between priors, Earth models and modified gravity constraints, which solves the ``non-linear error propagation'' problem.
For instance, we could fly two satellites at different altitudes at the same time to  break directly the degeneracy between the Newtonian and the Yukawa contributions to the spherical harmonic expansion of the gravitational field. Assuming that systematic errors are well-controlled, any difference between measurements done simultaneously would stem from modified gravity, which would naturally be accounted for in the underlying model.

To be complete, we should mention that a better way to beat the non-linear error propagation problem, instead of iterating between priors and updated models and constraints, would be to use Bayesian analysis, where the (posterior) probability density function of a set of parameters $\theta$ is computed from a data set $d$ as \cite{trotta08}
\begin{equation}
p(\theta|d) = \frac{p(d|\theta)p(\theta)}{p(d)},
\end{equation}
where $p(d|\theta)$ is the likelihood to have the current data set given the model parameters, $p(\theta)$ is the prior on the parameters, which encompasses our {\it a priori} knowledge of the model, and $p(d) = \int p(d|\theta)p(\theta)$ is a normalization constant.
This frame should be advantageously used to  relate efficiently geodesy and modified gravity experiments, non-linear problems being properly embedded in the (Bayesian) prior.
Techniques such as Markov Chain Monte Carlo regression could then be used to provide robust estimates of the Earth (Newtonian) gravitational field and of modified gravity.

Let us add a word of caution here. Although any current tension between gravitational field models coming from satellites at different altitudes may be explained as underestimated statistical errors, uncontrolled systematics (e.g. the time evolution of the shape of the Earth makes it particularly difficult to compare data obtained at different epochs) or as hints for modified gravity, it is expected that adding parameters to the gravity model will improve the fits and may relax the tension. However, this will not mean that the new model is better. Only model comparison techniques will then allow us to decide whether adding parameters to the model is relevant. A rich literature on model comparison in the closely related field of cosmology is available, that can serve as an introduction to the topic (e.g. \cite{trotta08, liddle04, seehars16, raveri18}).

We shall close this discussion by noting that we only discussed explicitly Earth-model-dependent constraints of modified gravity. For completeness, we briefly mention that some tests of (modified) gravity do not require any explicit Earth model. This is for instance the case of experiments that aim to look for a model-independent deviation to Newtonian gravity or GR and only need a gravity model as provided by gravity surveys to correct for systematics, with no explicit link to the real Earth geometry. For example, MICROSCOPE is sensitive to the Earth gravity gradient (GGT) \cite{touboul17}, which is therefore corrected for with published ITSG-GRACE14s spherical harmonic coefficients \cite{mayer06}. No error nor bias from any Earth model can thereby enter in the search for a violation of the equivalence principle. However, we warn that if a Yukawa interaction is present, then its effect at the MICROSCOPE altitude should not be the same than that at the altitude where the gravity model was measured by GRACE, potentially creating a bias in the GGT correction. However, from the orders of magnitude derived in Sect. \ref{sect_ofm}, we expect this possible bias to be negligible. On the opposite, the constraints on the Yukawa interaction estimated from the first MICROSCOPE results \cite{berge18} rely on an explicit model of the Earth; its impact will be assessed in a future work.

Finally, where possible, the most promising way to go beyond limitations from gravity surveys performed at different altitudes and from imperfect Earth models may be to embark a gravitational field measurement device onboard any satellite mission that aims to test modified gravity. For example, would a gradiometer surround the MICROSCOPE instrument, it could directly measure the actual GGT affecting the measurement,  which could then be corrected for without relying on any external gravity model. However, we do not see how to go pass the limitations from our imperfect knowledge of the Earth model in tests that are explicitly model-dependent (e.g. the expected Yukawa interaction-induced equivalence principle violation explicitly depends on the Earth physical characteristics --and not only its local gravitational field). An in-depth analysis of those limitations will be done in a future work.

\section{Order of magnitude estimates: homogeneous ellipsoidal Earth model} \label{sect_ofm}

In this section, we provide order-of-magnitude estimates of the impact that the imperfect knowledge of the shape of the Earth and a Yukawa interaction have on each other, as applications of the discussion in Sect. \ref{sect_impact}.
Without loss of generality, we consider a very simple Earth model, where the Earth is a rotationally symmetric, homogeneous ellipsoid. We can therefore use the results of Sect. \ref{sect_emodel}, with $N=1$.
We assume numerical values listed in Table \ref{tab_values}.
We should note that our model's flattening is not equal to the actual measured one: we chose it in order to recover the mass and $J_2$ measured for the actual Earth, despite having an overly simple Earth model.

\begin{table}
\caption{\label{tab_values}Homogeneous Earth model parameters: equatorial radius $R_\Earth$, density $\rho$ and (inverse) flatness $1/f$.}
\begin{indented}
\item[]\begin{tabular}{@{}lll}
\br
$R_\Earth$ & $\rho$ & 1/$f$  \\
\mr
 6378.1 km  &  $5.51\times 10^6$ g/m$^3$  & $370 \pm 10$ \\
\br
\end{tabular}
\end{indented}
\end{table}

\subsection{Impact of the Yukawa interaction on the measured Earth gravitational field} \label{sect_impact_cn0}

\subsubsection{Impact on the quadrupole}

We start with order-of-magnitude estimates of the contribution of the Yukawa interaction in the bias and variance of the Newtonian estimator of the $y_{20}$ coefficients, as an application of the discussion in Sect. \ref{ssect_h1}.

Using Eq. (\ref{eq_y20_homogeneous}) for the $y_{20}$ coefficient of a homogeneous Earth, the $\hat{y}_{20}^N$ estimator of Eq. (\ref{eq_ylm_estimator}) becomes
\begin{equation}
\hat{y}_{20}^N = \hat{y}_{20}(r) + \frac{2 \tilde{f} \tilde{\alpha} k(r,\lambda,R_\Earth)}{5\sqrt{5}(1-\tilde{f})}
\end{equation}
where $k(r, \lambda, R_\Earth) = 5{\rm e}^{-r/\lambda} \kappa\left(\frac{r}{\lambda} \right) \Phi_2\left( \frac{R_\Earth}{\lambda} \right)$ and where we assume that we experimentally measured $\hat{y}_{20}(r)$. As before, the tilde symbols represents priors.

The expected value and variance of this estimator, derived from Eqs. (\ref{eq_ylm_ev})-(\ref{eq_ylm_var}) give
\begin{equation} \label{eq_y20_ev}
\E(\hat{y}_{20}^N) = y_{20}^N + \frac{2 f k(r,\lambda,R_\Earth)}{5\sqrt{5}(1-f))} \delta\alpha
\end{equation}
where we assume that the measurement and flattening model are unbiased but the prior on $\alpha$ is biased ($\E(\tilde{\alpha})=\alpha+\delta\alpha$), and
\begin{equation} \label{eq_y20_var}
\sigma_{y20N}^2 = \sigma_{y20}^2 + \frac{4 k^2(r,\lambda,R_\Earth)}{125(1-\tilde{f})} \left[ \frac{\tilde{f}\tilde{\alpha}}{(1-\tilde{f})^2} \sigma_{\tilde{f}}^2 + \sigma_{\tilde{\alpha}}^2 \right].
\end{equation}

We can note that when ignoring the possibility of a non-zero Yukawa interaction, the bias in Eq. (\ref{eq_y20_ev}) is just the Yukawa contribution to the $y_{20}(r)$ coefficient (Eq. \ref{eq_y20_homogeneous}).

\begin{figure}
\includegraphics[width=0.55\textwidth]{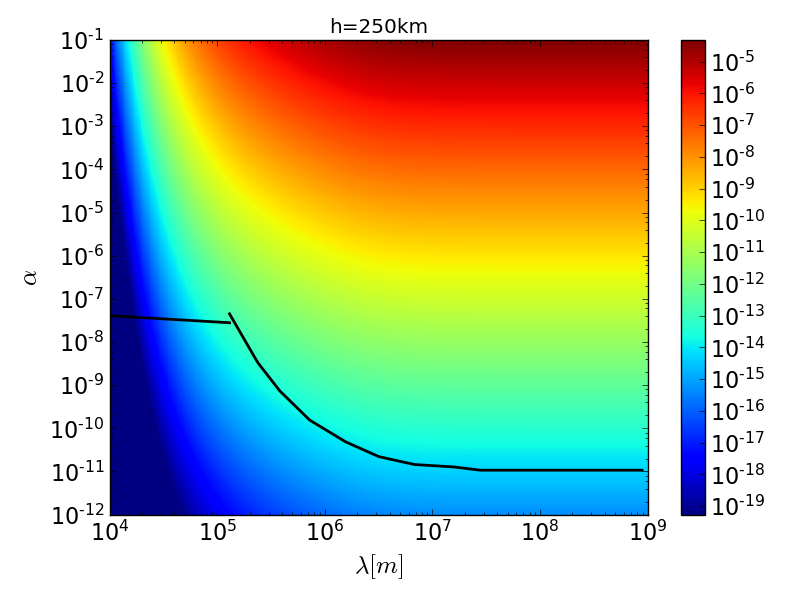}
\includegraphics[width=0.55\textwidth]{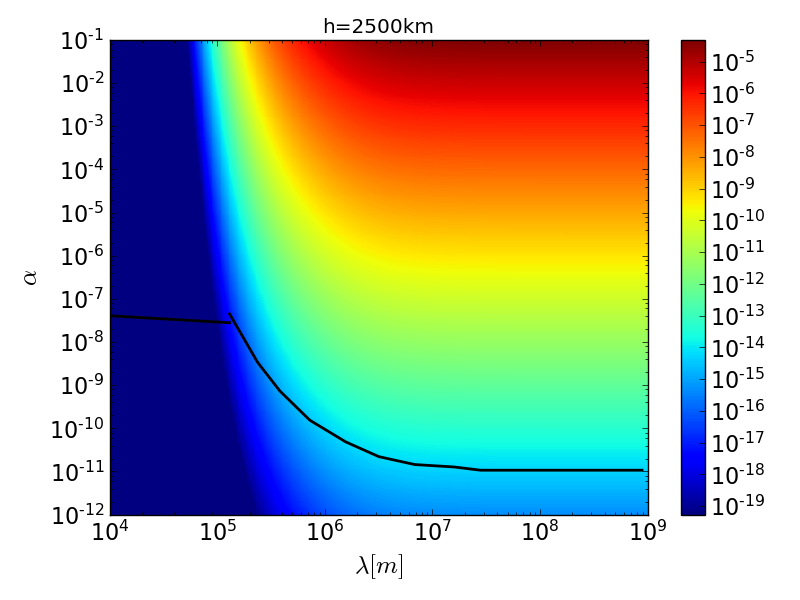}
\caption{Bias on the estimated $y_{20}^N$ Newtonian zonal term from a Yukawa deviation for a homogeneous Earth (Eq. \ref{eq_y20_ev}) when incorrectly assuming $\alpha=0$ if a non-zero Yukawa interaction actually exists, in the $\alpha-\lambda$ plane, for two experiments orbiting the Earth at different altitudes: 250km (like GOCE) and 2500 km. In this case, the bias is just the Yukawa contribution to the $y_{20}(r)$ coefficient (Eq. \ref{eq_y20_homogeneous}).
The black contour shows the best existing exclusion constraints on a Yukawa interaction \cite{berge18,kolosnitsyn04}.}
\label{fig_deltac202d}       
\end{figure}

Fig. \ref{fig_deltac202d} shows the bias on the estimated $\hat{y}_{20}^N$ as given by Eq. (\ref{eq_y20_ev}), when (incorrectly) assuming $\alpha=0$, for a low-earth orbit experiment (altitude of GOCE --left panel) and a hypothetical mid-earth orbit (2500 km --right panel) in the $\alpha-\lambda$ plane. The black line shows the current best constraints on the existence of a Yukawa interaction \cite{berge18,kolosnitsyn04}: the region of the plane above the line is excluded by previous experiments. It is clear that the effect of a given ($\alpha$, $\lambda$) pair affects the measurement of $y_{20}$ differently depending on the altitude, due to the exponential dependence of the Yukawa interaction. For instance, ($\alpha$, $\lambda$) $\approx$ ($2\times10^{-8}$, $1.2\times10^{5}$m), i.e.  for $\delta\alpha= 2\times10^{-8}$,  brings a bias of about $10^{-13}$ for an experiment at the GOCE altitude, while it barely affects an experiment at 2500 km ($\delta y_{20}\approx 10^{-16}$).
The exact value for a 250-km and 500-km altitude satellites is given in Table \ref{tab_deltaC}.

Assuming a prior $\tilde{\alpha}=0\pm\sigma_{\tilde\alpha}$, and still assuming that the model of the flattening is unbiased, Eq. (\ref{eq_y20_var}) shows that the increase in the measured $y_{20}^N$'s uncertainty is equal to the maximum bias that can be brought by allowed values for the Yukawa parameters. This once again shows that choosing a good prior on $\alpha$ helps to minimise the bias on $y_{20}^N$ (at the price of increasing its error bar).

\begin{table}
\caption{\label{tab_deltaC}Bias on the first few estimated Newtonian zonal terms from a Yukawa deviation for a rotationally symmetric, homogeneous ellipsoidal Earth, when incorrectly assuming $\alpha=0$ if a non-zero Yukawa interaction actually exists with ($\alpha$, $\lambda$) = ($2\times10^{-8}$, $1.2\times10^{5}$m), at an altitude of 250 km (GOCE \cite{pail11}) and 500 km (GRACE \cite{mayer06}).}
\begin{indented}
\item[]\begin{tabular}{@{}llllll}
\br
 & Yukawa bias & Tabulated value & Tabulated \\
 & (rms increase &  & uncertainty \\
  & -- ${\rm E}(\hat{\alpha})=0$)  \\
\mr
GOCE \\
\mr
$y_{20}$ & $7.4\times10^{-14}$ &  $-4.84165304245\times10^{-4}$ & $5.423\times10^{-12}$ \\
$y_{40}$ & $1.3\times10^{-15}$ & $5.39950509\times10^{-7}$ & $2.758\times10^{-12}$ \\
$y_{60}$ & $2.5\times10^{-15}$ &  $-1.49979681\times10^{-7}$ & $3.556\times10^{-12}$ \\
$y_{80}$ & $4.1\times10^{-15}$ & $4.9448989\times10^{-8}$ & $3.972\times10^{-12}$ \\
\mr
GRACE \\
\mr
$y_{20}$ & $1.0\times10^{-14}$ &  $-4.84169283673\times10^{-4}$ & $1.577\times10^{-12}$ \\
$y_{40}$ & $1.8\times10^{-16}$ &  $5.39993370\times10^{-7}$ & $3.35\times10^{-13}$ \\
$y_{60}$ & $3.8\times10^{-16}$ & $-1.49974614\times10^{-7}$ & $1.88\times10^{-13}$ \\
$y_{80}$ & $6.7\times10^{-16}$ & $4.9477947\times10^{-8}$ & $1.35\times10^{-13}$ \\
\br
\end{tabular}
\end{indented}
\end{table}

\subsubsection{Impact on higher zonal terms}

Table \ref{tab_deltaC} lists the expected deviations for $y_{n0}$ ($n=2,4,6,8$) due to a Yukawa interaction, for a rotationally symmetric, homogeneous ellipsoidal Earth, at altitudes of 250 km and 500 km, and compares them with current uncertainties on the measured coefficients for GOCE-only and GRACE-only gravitational field models \cite{pail11, mayer06}. Those numbers are normalised such that the Newtonian contributions correspond to the measurements for the actual Earth, to account for our oversimplified Earth model.
The first column gives the expected bias from a Yukawa interaction with ($\alpha$, $\lambda$) = ($2\times10^{-8}$, $1.2\times10^{5}$m), or equivalently the increase in rms for ($\alpha$, $\lambda$) = ($0\pm2\times10^{-8}$, $1.2\times10^{5}$m); the third and fourth columns give up-to-date tabulated values.

The results listed in the table show that current space geodesy missions, which fly higher than a few hundred kilometers, are immune to a Yukawa interaction (as currently constrained by other experiments). Currently allowed values of Yukawa parameters only marginally affect the measurement of the Newtonian spherical harmonics: the expected bias (equivalently, uncertainty increase would a Yukawa interaction be absent, but our imperfect knowledge about it considered) is between two and three orders of magnitude smaller than the current errors on the first few zonal terms.
Nevertheless, should the measurement errors be decreased by two orders of magnitude (even for high-altitude satellites), care should be taken to include the Yukawa interaction in the model.

\subsection{Impact of the Earth geometry and mass distribution on the constraints on Yukawa parameters} \label{sect_impact_yukawa}

We consider the impact of our imperfect knowledge of the Earth shape and compute an order of magnitude estimate of the level of error that we may expect on the estimation of $\alpha$. In this section, we consider that we constrain $\alpha$ for fixed $\lambda$ (then the $\alpha-\lambda$ plane can be constrained by binning it along $\lambda$) and use the estimator (\ref{eq_alpha_estimator}).

We keep the same Earth model (Table \ref{tab_values}), where we assume some error on the flattening ($\delta f/f=0.027$). In the case of an homogeneous Earth, the estimator's expected value is given by
\begin{equation}
\E(\hat{\alpha}) = \alpha - \frac{5\sqrt{5}}{2f [k(r_1,\lambda,R_\Earth) - k(r_2,\lambda,R_\Earth)]} \frac{\delta f}{f} (y_{20}(r_1) - y_{20}(r_2))
\end{equation}
where we assumed that the $\hat{y}_{20}$ measurements are unbiased, and that the model of the flattening is biased by $\delta f$.
Assuming that $\alpha = 2\times10^{-8}$ and $\lambda = 1.2\times10^{5}$m, and that the satellites orbit the Earth at 250 km and 2500 km (which allow for the larger difference $y_{20}(r_1) - y_{20}(r_2)$ in the allowed region of the ($\alpha$, $\lambda$) plane --see Fig. \ref{fig_deltac202d}), we find a 40\% bias $\delta \alpha = 8\times10^{-9}$ on the estimation of $\alpha$. This is a significant bias, that may point to a close limitation due to our knowledge of the Earth.
However, our homogeneous Earth model is deliberately simplistic and implies a large error on the flattening. Since the bias on $\alpha$ scales linearly with the relative uncertainty on the flattening, we can expect that better Earth models (e.g. 2-layer models), with smaller error on the flattening, will have a less significant bias on the constraints on $\alpha$.

\begin{figure}
\begin{center}
\includegraphics[width=0.55\textwidth]{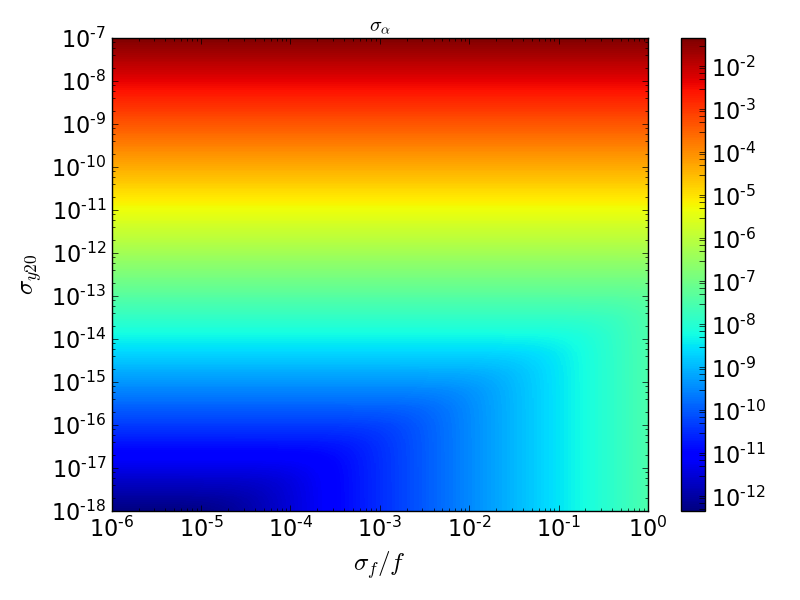}
\caption{Contribution to the $\hat{\alpha}$ estimator variance of the relative error on the modeled Earth flattening and of measurement error on $y_{20}$, for $\lambda=1.2\times10^5$m.}
\label{fig_sigmaJ2}       
\end{center}
\end{figure}

The uncertainty on the $\hat{\alpha}$ estimator is given (at first order) by
\begin{equation}
\sigma_\alpha = \frac{5\sqrt{5}(1-f)}{2f[k(r_1,\lambda,R_\Earth) - k(r_2,\lambda,R_\Earth)]} \sqrt{\frac{\hat{y}_{20}(r_1) - \hat{y}_{20}(r_2)}{(1-f)^2} \frac{\sigma_f^2}{f^2} + 2 \sigma_{y20}^2}
\end{equation}
where we assumed $\sigma_{y20}^2(r_1) = \sigma_{y20}^2(r_2)$ and ignored any bias on the flattening, but consider some uncertainty $\sigma_f$ on it.
As discussed in Sect. \ref{ssect_h2}, the uncertainty on the $\alpha$ estimator has contributions from the measurement errors and from the uncertainty on the Earth model. Fig. \ref{fig_sigmaJ2} compares those two contributions. It should be noted that in our simple example, if we assume a percent error on the flattening, the $y_{20}$ measurement errors dominate down to $\sigma_{y20} \approx 10^{-15}$. As soon as gravity surveys reach a better precision, then the Earth model will limit experiments aiming to constraints a Yukawa interaction.

Since the relative error on the mass of the Earth scales linearly with the relative error on the flattening, Fig. \ref{fig_sigmaJ2} can be used to confirm the crude estimate that we made in Sect. \ref{ssect_h2}: with $\sigma_f / f \approx \sigma_{M_\Earth} / M_\Earth \approx 10^{-4}$, this uncertainty will dominate over the $y_{20}$ measurement errors as soon as the latter are better than $10^{-17}$ (in the case presently under consideration, where ($\alpha$, $\lambda$) = ($2\times10^{-8}$, $1.2\times10^{5}$m)).

Although the numbers given in this section are meant for rough order-of-magnitude estimates, they show that current experiments are not yet limited by our ability to reliably model the Earth.

\subsection{Impact of a Yukawa interaction on orbital dynamics} \label{ssect_ofm_pert}

\begin{figure}
\begin{center}
\includegraphics[width=0.7\textwidth]{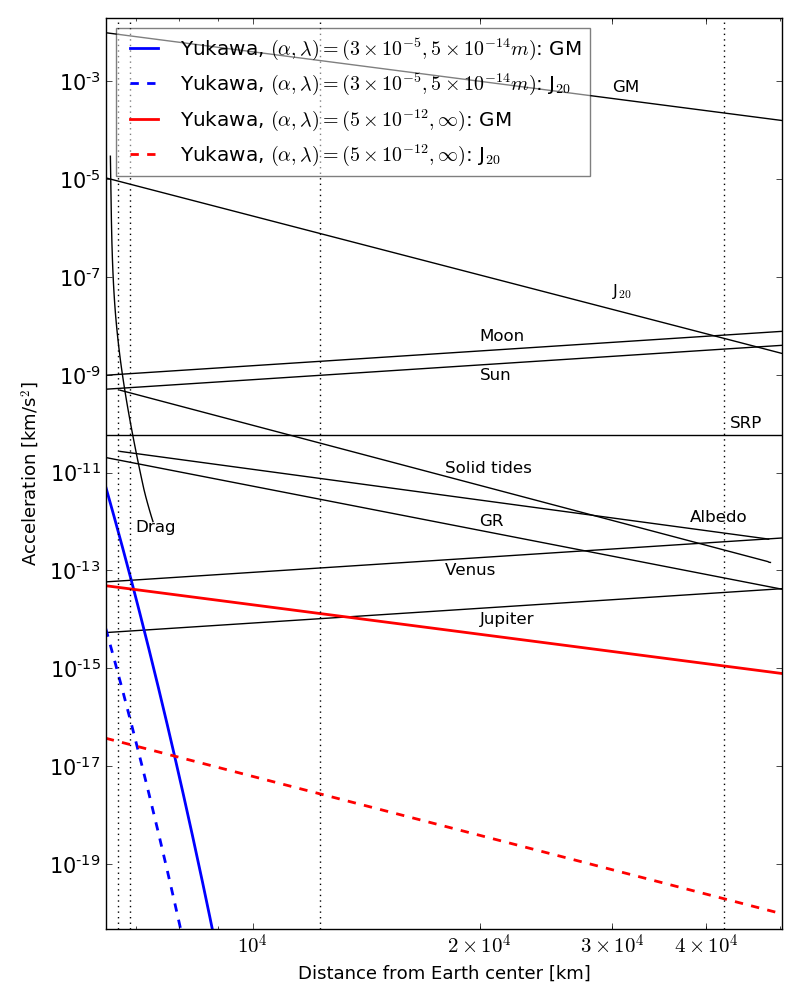}
\caption{Comparison of Yukawa interaction perturbation with usual gravitational and non-gravitational perturbations, for satellites up to geostationary altitude. Black solid lines represent usual perturbations, as can be computed e.g. from Ref.  \cite{montenbruck} (see main text). Colored lines show the Yukawa perturbation for two different allowed configurations: short-range, relatively strong interaction (blue) and long-range, weak interaction (red); solid lines show the acceleration of the Yukawa interaction monopole, and dashed lines show the acceleration due to the Yukawa interaction quadrupole. Dotted lines show the altitude of GOCE, GRACE, LAGEOS and geostationary satellites from left to right.}
\label{fig_apertall}       
\end{center}
\end{figure}

\begin{figure}
\begin{center}
\includegraphics[width=0.7\textwidth]{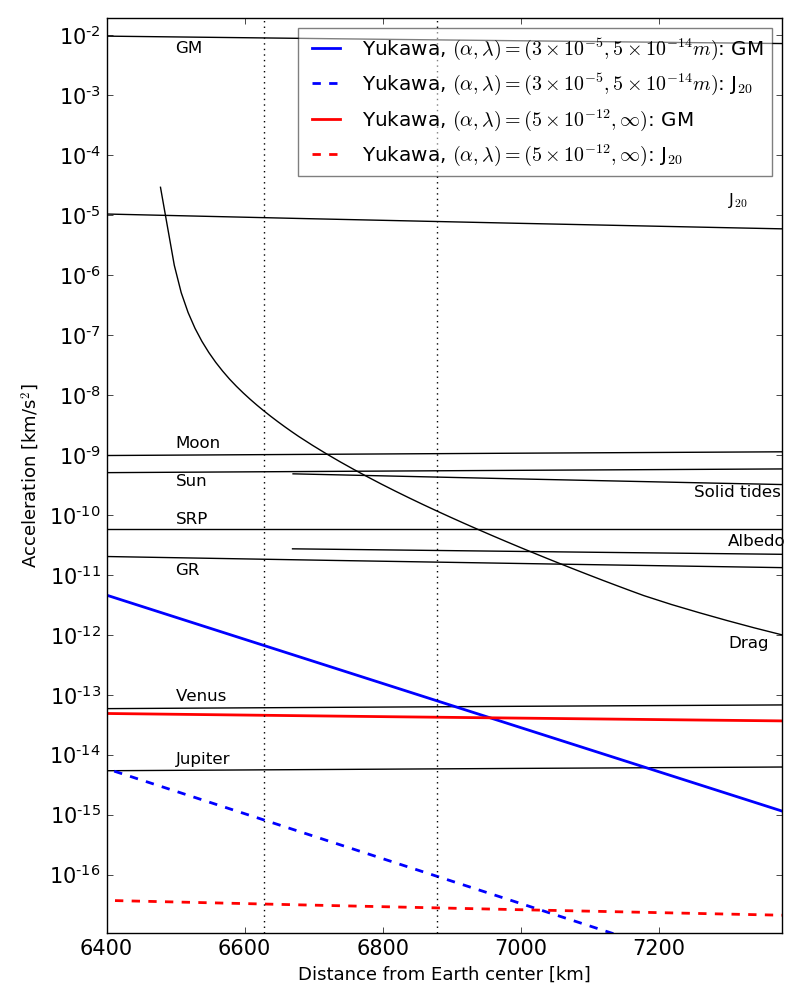}
\caption{Same as Fig. \ref{fig_apertall}, for low-Earth orbits.}
\label{fig_apertleo}       
\end{center}
\end{figure}

We now quantify the perturbing accelerations created by a Yukawa interaction on an orbiting satellite (see Sect. \ref{ssect_exg}). Figs. \ref{fig_apertall} and \ref{fig_apertleo} compare the Yukawa monopole (i.e. the term in $z_{00}(r)$ in ${\bf g_\parallel}$,  and quadrupole accelerations (i.e. the term in $J_{20}(r)$ in ${\bf g_\perp}$),  to other usual gravitational and non-gravitational accelerations. Their effect is shown for altitudes up to the geostationary altitude in Fig. \ref{fig_apertall}, while Fig. \ref{fig_apertleo} zooms on low-earth orbits. The blue lines correspond to our fiducial ($\alpha$, $\lambda$) = ($2\times10^{-8}$, $1.2\times10^{5}$m) model, and the red lines represent a long-range Yukawa interaction ($\alpha$, $\lambda$) = ($5\times10^{-12}$, $\infty$), as still allowed by experiments. In each case, the solid line corresponds to the monopole acceleration and the dashed line shows the quadrupole acceleration.

The other lines show the acceleration of the Earth Newtonian monopole (GM), and several gravitational (Newtonian Earth quadrupole --$J_{20}$--, gravitational pull of the Moon, Sun, Venus and Jupiter, relativistic effects --GR--, Earth tides) and non-gravitational (solar radiation pressure --SRP--, atmospheric drag, Earth albedo) perturbations. We followed Ref. \cite{montenbruck} to compute those perturbations. The line showing the atmospheric drag is based on an upper limit of the atmospheric density, and therefore shows the maximum drag expected. The vertical dotted lines show the altitude of GOCE, GRACE, LAGEOS and geostationary satellites from left to right.

A long-ranged Yukawa interaction is largely subdominant for altitudes higher than a few thousands kilometers; below that, its perturbation is of the order of those of Venus and Jupiter. In particular, the perturbation due to the coupling between the Earth's quadrupole and a long-ranged Yukawa interaction is several orders of magnitude lower than the perturbation caused by Jupiter.

Perturbations caused by a mid-ranged Yukawa interaction (as still allowed by experiments) fall off quickly with the altitude, so that they are ever more subdominant than a long-ranged Yukawa interaction for satellites orbiting the Earth higher than 500 km. However, they may have an  impact similar to that of relativistic effects on low-earth satellites; the quadrupole acceleration, although less significant, can be of the same order as the perturbations caused by Venus and Jupiter.

Finally, Figs. \ref{fig_apertall} and \ref{fig_apertleo} clearly show the strong radial dependence of the Yukawa interaction that we mentioned throughout this paper. It means that satellites like GOCE and GRACE are not affected in the same way by a Yukawa interaction, although other perturbations (leaving apart the atmospheric drag) impact both of them in a similar manner. This confirms the possibility to use two such satellites to constrain a Yukawa interaction in low-earth orbit, as we have sketched in Sect. \ref{sect_impact_yukawa}, or directly through the comparison of their dynamics. This can be done by solving Lagrange-Gauss equations, which we will present in a future work.

\section{Conclusion}

We have investigated the entanglement between the shape of the Earth and modified gravity. Describing deviations to Newtonian gravity with a Yukawa interaction, we showed that the Earth gravitational field potential can still be expanded in spherical harmonics, just like in the pure Newtonian realm.
We derived explicit expressions for the spherical harmonic coefficients, that we used to compute the (modified) gravity acceleration. We finally considered the Lagrange-Gauss equations, that describe the effect of a perturbing force on a satellite's orbital dynamics, in the case where the Yukawa interaction is sourced by the complex shape of the Earth.
To perform those calculations, we introduced a new method to compute a multipolar decomposition of the gravity acceleration with spin-weighted spherical harmonics, which greatly simplifies the required algebra.

We showed that although formally the coefficients of the spherical harmonic expansion keep the same form as in the Newtonian case, they acquire a new meaning and are not universal to the Earth system anymore, since they become explicitly dependent on the distance from the centre of the Earth.
Consequently, the gravitational acceleration and the perturbing force due to the shape of the Earth also acquire a new radial dependence.

This behavior has many implications both in geodesy and in modified gravity experiments:
\begin{itemize}
\item in presence of a non-zero Yukawa interaction, measurements of the Earth gravitational field performed at different altitudes inevitably provide inconsistent results (up to measurement errors). 
\item in presence of a non-zero Yukawa interaction, using a Newtonian gravity model to map the Earth mass distribution by inverting the spherical harmonic coefficients measured for the gravitational field is prone to be biased; using a prior on modified gravity, considered as a systematic error, should help to minimise the bias, although the uncertainty on the mass distribution estimator will increase.
\item Earth-model-dependent measurements of a Yukawa interaction are inevitably affected by any bias or uncertainty on the Earth model (e.g. coming from geodesy data). Model-independent estimators might be constructed but require that gravity surveys go beyond the implicit assumption that the underlying field is Newtonian.
\item even experiments that rely only on the measured Earth gravitational field (with no need to detail its source) are prone to errors if they are performed at an altitude different from that where the gravitational field was measured.
\end{itemize}

We proposed to combine gravitational surveys to define a new estimator of the Yukawa interaction strength $\alpha$. Taking advantage of the radial dependence of the spherical harmonic coefficients in presence of a Yukawa interaction, we can simply take the difference of the values of a given coefficient as measured by two satellites at different altitude. We discussed the limitations caused by our imperfect knowledge of the Earth. Despite a significant bias in $\alpha$ if the model of the Earth is too simplistic, we found that we can increase the instrumental precision by several orders of magnitude before being limited by our knowledge of the Earth. However, we restrained from deriving new constraints on the Yukawa interaction from the strong tension in the $J_2$ zonal term as measured by GOCE and GRACE, since we find it dubious and its most probable cause is underestimated measurement errors.

Although the limitations listed above seem profound, we showed that they are subdominant compared to other usual gravitational and non-gravitational perturbations. We based our conclusion on order-of-magnitude estimates using a simple Earth model and taking into account those values of the Yukawa interaction that are still allowed by experiments but that give the strongest effects. For instance, the strength of the perturbation imparted by the coupling of the Earth quadrupole with a Yukawa interaction on a satellite is smaller than that due to Jupiter. Very-low-altitude satellites could be affected by a mid-range, still undetected Yukawa interaction, at the level of usual relativistic effects. Thus, it is from low-altitude experiments that it seems most likely to improve our knowledge about a possible Yukawa interaction, provided that the atmospheric drag can be correctly taken into account (e.g. through a drag-free system).

We can therefore expect that although we should rigorously take into account the complex shape of the Earth when constraining modified gravity in orbit, especially for experiments performed in a low-Earth orbit, considering the Earth as a sphere remains a very good approximation for high-altitude satellites. Nevertheless, it would be sound to gather geodesy and modified gravity to minimise any modeling limitation. This can be done by performing geodesy experiments with modified gravity in mind (i.e. using a beyond-Newton gravity model), or even by designing experiments aiming to measure the shape of the Earth and modified gravity simultaneously.

\appendix
\section{Proof of Eq. (\ref{eq_H})} \label{app_phi}

We wish to compute
\begin{equation}
H \equiv \int_a^b  x^{n+2} \left( \frac{\lambda}{R_\Earth x}\right)^{n+\frac{1}{2}} I_{n+\frac{1}{2}} \left( \frac{R_\Earth x}{\lambda}\right) {\rm d}x
\end{equation}
Let us first introduce the new variables $k=R_\Earth/\lambda$ and $y=kx$, such that
\begin{equation}
H = k^{-n-3} \int_{ak}^{bk} y^{3/2} I_{n+\frac{1}{2}}(y) {\rm d}y.
\end{equation}
We then define
\begin{equation}
\phi_\ell(x,k) = k^{-n-3} \int_0^{kx} y^{3/2} I_{n+\frac{1}{2}}(y) {\rm d}y
\end{equation}
such that $H = \phi_\ell(b,k)- \phi_\ell(a,k)$.
Using \cite{cuyt}
\begin{equation}
I_\alpha(x) = \frac{\left(\frac{x}{2}\right)^\alpha}{\Gamma(\alpha+1)} {}_0F_1(;\alpha+1;\frac{x^2}{4}),
\end{equation}
where ${}_0F_1()$ is the confluent hypergeometric limit function, and setting $u = \frac{y}{kx}$, we get
\begin{equation}
\phi_\ell(x,k) = \frac{2^{-n-\frac{1}{2}} x^{n+3}}{\Gamma\left(n+\frac{3}{2}\right)} \int_0^1 u^{n+2} {}_0F_1\left(;n+\frac{3}{2};\frac{k^2x^2}{4}u^2\right) {\rm d}u.
\end{equation}
An extra change of variable $v=u^2$ provides
\begin{equation}
\phi_\ell(x,k) = \frac{2^{-n-\frac{3}{2}}x^{n+3}}{\Gamma\left(n+\frac{3}{2}\right)} \int_0^1 v^\frac{n+1}{2} {}_0F_1\left(;n+\frac{3}{2};\frac{k^2x^2}{4} v\right) {\rm d}v.
\end{equation}
Finally, using \cite{cuyt}
\begin{multline}
{}_{A+1}F_{B+1}(a_1, \dots, a_A, c; b_1, \dots, b_B, d; z) = \\
\frac{\Gamma(d)}{\Gamma(c) \Gamma(d-c)} \int_0^1 t^{c-1}(1-t)^{(d-c-1)} {}_AF_B(a_1,\dots,a_A; b_1,\dots,b_B;tz) {\rm d}t,
\end{multline}
we obtain
\begin{equation}
\phi_\ell(x,k) = 2^{-n-\frac{3}{2}} x^{n+3} \frac{\Gamma\left(\frac{n+3}{2}\right)}{\Gamma\left(n+\frac{3}{2}\right) \Gamma\left(\frac{n+5}{2}\right)} {}_1F_2\left(\frac{n+3}{2}; n+\frac{3}{2}, \frac{n+5}{2}; \frac{k^2x^2}{4}\right),
\end{equation}
which proves Eq. (\ref{eq_H}).

\section{Form factors} \label{app_phi2}

This appendix discusses some aspects of the form factors introduced in Sect. \ref{sect_emodel} for a homogeneous, rotationally symmetric ellipsoid of flatness $f$ and equatorial radius $R_E$
\begin{eqnarray}
\Phi(x,f) &=& 3 \frac{x \cosh(x) - \sinh(x)}{x^3} - \frac{\sinh x}{x} f \\
\Phi_2(x) &=& 3 \frac{x \cosh(x) - \left(x^2/3+1\right)\sinh(x)}{x^5}.
\end{eqnarray}

\begin{figure}
\includegraphics[width=0.5\textwidth]{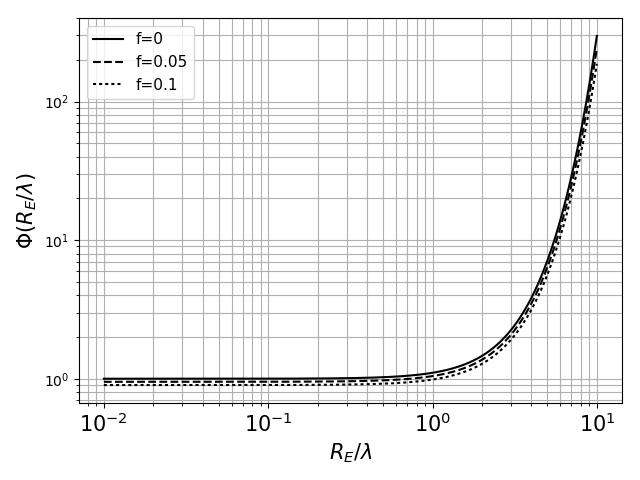}
\includegraphics[width=0.5\textwidth]{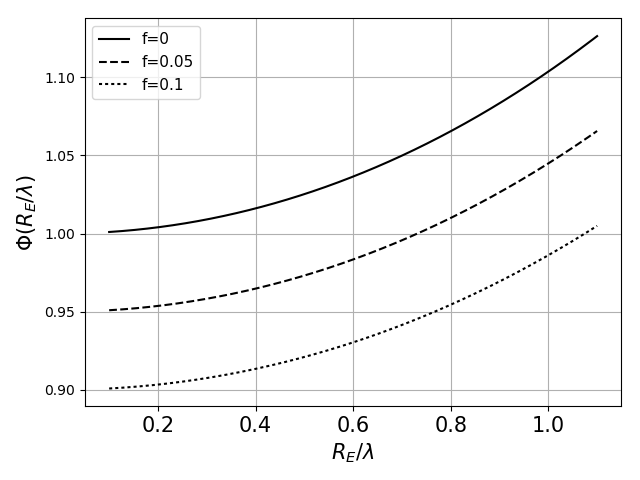}
\includegraphics[width=0.5\textwidth]{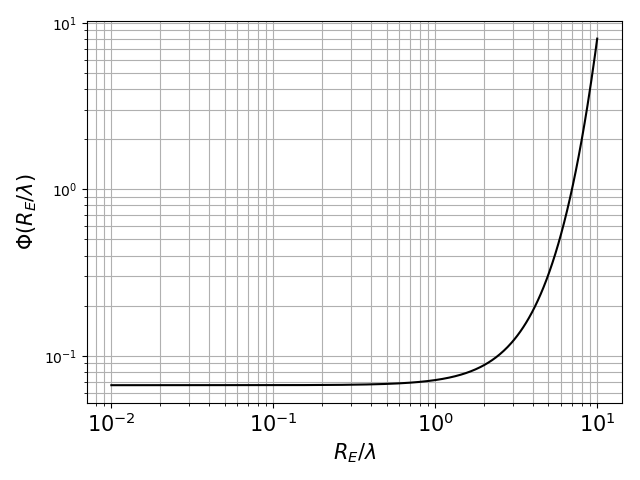}
\includegraphics[width=0.5\textwidth]{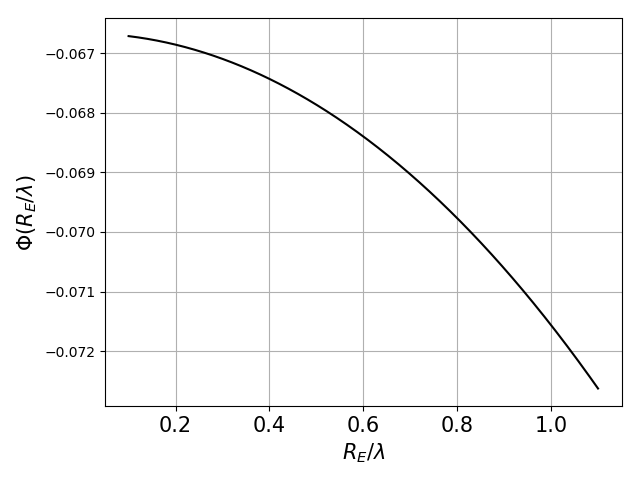}
\caption{Upper panels: form factor $\Phi(x,f)$, as a function of the ratio $R_\Earth/\lambda$ (left: logarithmic scale; right: zoom about $\lambda \approx R_\Earth$). Lower panels: form factor $\Phi_2(x)$, as a function of the ratio $R_\Earth/\lambda$ (left: logarithmic scale; right: zoom about $\lambda \approx R_\Earth$). Long-range Yukawa interaction set on the left of the plots, while short-range Yukawa interaction are on the right.}
\label{fig_Phi2}       
\end{figure}

They are shown in Fig. \ref{fig_Phi2}, as a function of $R_E/\lambda$. The upper panels show $\Phi(R_E/\lambda,f)$ for three different flatnesses; it is clear that the flatness introduces a linear offset (note that the flatnesses used in the figure are much higher than the actual flatness of the Earth). The lower panels show $\Phi_2(R_E/\lambda)$.
For long-range interactions ($R_E/\lambda \rightarrow 0$), both function tend to a finite limit: $\Phi(R_E/\lambda,f) \rightarrow 1-f$ and $\Phi_2(R_E/\lambda) \rightarrow -1/15$. In this case, the form factor does not play a role in the monopole acceleration (up to the flatness), but it limits the quadrupole acceleration.
Short-range interactions are more strongly affected by those form factors, highlighting the fact that Gauss theorem does not apply to a Yukawa interaction.
In particular, for $\lambda \sim 0.1 R_E$, the Yukawa monopole acceleration is boosted by 2 orders of magnitude, meaning that it does not scale naively as $\alpha g_{\rm Newton}$, but as $100 \alpha g_{\rm Newton}$. Therefore, correctly taking this form factor into account is important to get correct constraints on the Yukawa interaction.

\ack
We thank Phuong-Anh Huynh and Bernard Foulon for useful and stimulating discussions. We acknowledge the financial support of CNES through the APR program (``GMscope+'' project) and of the UnivEarthS Labex program at Sorbonne Paris Cit\'e (ANR-10-LABX-0023 and ANR-11-IDEX-0005-02). The work of JPU is made in the ILP LABEX (under reference ANR-10-LABX-63) was supported by French state funds managed by the ANR within the Investissements d'Avenir programme under reference ANR-11-IDEX-0004-02. This work is supported in part by the EU Horizon 2020 research and innovation programme under the Marie-Sklodowska grant No. 690575. This article is based upon work related to the COST Action CA15117 (CANTATA) supported by COST (European Cooperation in Science and Technology).

\section*{References}
\bibliographystyle{iopart-num}
\bibliography{yuk18}

\providecommand{\noopsort}[1]{}\providecommand{\singleletter}[1]{#1}%
\providecommand{\newblock}{}
\begin{thebibliography}{10}
\expandafter\ifx\csname url\endcsname\relax
  \def\url#1{{\tt #1}}\fi
\expandafter\ifx\csname urlprefix\endcsname\relax\def\urlprefix{URL }\fi
\providecommand{\eprint}[2][]{\url{#2}}

\bibitem{Will:1993ns}
Will C~M 1993 {\em {Theory and experiment in gravitational physics}\/} ISBN
  9780521439732

\bibitem{will14}
{Will} C~M 2014 {\em Living Reviews in Relativity\/} {\bf 17} 4
  (\textit{Preprint} \eprint{1403.7377})

\bibitem{adelberger03}
{Adelberger} E~G, {Heckel} B~R and {Nelson} A~E 2003 {\em Annual Review of
  Nuclear and Particle Science\/} {\bf 53} 77--121 (\textit{Preprint}
  \eprint{hep-ph/0307284})

\bibitem{clifton12}
{Clifton} T, {Ferreira} P~G, {Padilla} A and {Skordis} C 2012 {\em Physics
  Reports\/} {\bf 513} 1--189 (\textit{Preprint} \eprint{1106.2476})

\bibitem{joyce15}
{Joyce} A, {Jain} B, {Khoury} J and {Trodden} M 2015 {\em Physics Reports\/}
  {\bf 568} 1--98 (\textit{Preprint} \eprint{1407.0059})

\bibitem{Uzan:2000mz}
Uzan J~P and Bernardeau F 2001 {\em Phys. Rev.\/} {\bf D64} 083004
  (\textit{Preprint} \eprint{hep-ph/0012011})

\bibitem{Uzan:2003zq}
Uzan J~P 2003 {\em Annales Henri Poincare\/} {\bf 4} S347--S369

\bibitem{Uzan:2006mf}
Uzan J~P 2007 {\em Gen. Rel. Grav.\/} {\bf 39} 307--342 (\textit{Preprint}
  \eprint{astro-ph/0605313})

\bibitem{Uzan:2010ri}
Uzan J~P 2010 {\em Gen. Rel. Grav.\/} {\bf 42} 2219--2246 (\textit{Preprint}
  \eprint{0908.2243})

\bibitem{Jain:2013wgs}
Jain B {\em et~al.\/} 2013  (\textit{Preprint} \eprint{1309.5389})

\bibitem{jain10}
{Jain} B and {Khoury} J 2010 {\em Annals of Physics\/} {\bf 325} 1479--1516
  (\textit{Preprint} \eprint{1004.3294})

\bibitem{safronova17}
{Safronova} M~S, {Budker} D, {DeMille} D, {Kimball} D~F~J, {Derevianko} A and
  {Clark} C~W 2017 {\em ArXiv e-prints\/} (\textit{Preprint}
  \eprint{1710.01833})

\bibitem{delva17}
{Delva} P, {Hees} A and {Wolf} P 2017 {\em Space Science Reviews\/} {\bf 212}
  1385--1421

\bibitem{everitt11}
{Everitt} C~W~F, {Debra} D~B, {Parkinson} B~W, {Turneaure} J~P, {Conklin} J~W,
  {Heifetz} M~I, {Keiser} G~M, {Silbergleit} A~S, {Holmes} T, {Kolodziejczak}
  J, {Al-Meshari} M, {Mester} J~C, {Muhlfelder} B, {Solomonik} V~G, {Stahl} K,
  {Worden} Jr P~W, {Bencze} W, {Buchman} S, {Clarke} B, {Al-Jadaan} A,
  {Al-Jibreen} H, {Li} J, {Lipa} J~A, {Lockhart} J~M, {Al-Suwaidan} B, {Taber}
  M and {Wang} S 2011 {\em Physical Review Letters\/} {\bf 106} 221101
  (\textit{Preprint} \eprint{1105.3456})

\bibitem{ciufolini13a}
{Ciufolini} I, {Paolozzi} A, {Koenig} R, {Pavlis} E~C, {Ries} J, {Matzner} R,
  {Gurzadyan} V, {Penrose} R, {Sindoni} G and {Paris} C 2013 {\em Nuclear
  Physics B Proceedings Supplements\/} {\bf 243} 180--193 (\textit{Preprint}
  \eprint{1309.1699})

\bibitem{ciufolini13b}
{Ciufolini} I, {Moreno Monge} B, {Paolozzi} A, {Koenig} R, {Sindoni} G,
  {Michalak} G and {Pavlis} E~C 2013 {\em Classical and Quantum Gravity\/} {\bf
  30} 235009 (\textit{Preprint} \eprint{1310.2601})

\bibitem{iorio02a}
{Iorio} L, {Ciufolini} I and {Pavlis} E~C 2002 {\em Classical and Quantum
  Gravity\/} {\bf 19} 4301--4309 (\textit{Preprint} \eprint{gr-qc/0103088})

\bibitem{lucchesi03}
{Lucchesi} D~M 2003 {\em Physics Letters A\/} {\bf 318} 234--240

\bibitem{lucchesi10}
{Lucchesi} D~M and {Peron} R 2010 {\em Physical Review Letters\/} {\bf 105}
  231103 (\textit{Preprint} \eprint{1106.2905})

\bibitem{lucchesi14}
{Lucchesi} D~M and {Peron} R 2014 {\em Phys. Rev. D\/} {\bf 89} 082002

\bibitem{li14}
{Li} Z~W, {Yuan} S~F, {Lu} C and {Xie} Y 2014 {\em Research in Astronomy and
  Astrophysics\/} {\bf 14} 139-143

\bibitem{bertotti03}
{Bertotti} B, {Iess} L and {Tortora} P 2003 {\em Nature\/} {\bf 425} 374--376

\bibitem{williams04}
{Williams} J~G, {Turyshev} S~G and {Boggs} D~H 2004 {\em Phys. Rev. Lett.\/}
  {\bf 93} 261101 (\textit{Preprint} \eprint{gr-qc/0411113})

\bibitem{schlamminger08}
{Schlamminger} S, {Choi} K~Y, {Wagner} T~A, {Gundlach} J~H and {Adelberger} E~G
  2008 {\em Physical Review Letters\/} {\bf 100} 041101 (\textit{Preprint}
  \eprint{0712.0607})

\bibitem{wagner12}
{Wagner} T~A, {Schlamminger} S, {Gundlach} J~H and {Adelberger} E~G 2012 {\em
  Class. Quant. Grav.\/} {\bf 29} 184002 (\textit{Preprint} \eprint{1207.2442})

\bibitem{touboul17}
{Touboul} P, {M{\'e}tris} G, {Rodrigues} M, {Andr{\'e}} Y, {Baghi} Q,
  {Berg{\'e}} J, {Boulanger} D, {Bremer} S, {Carle} P, {Chhun} R, {Christophe}
  B, {Cipolla} V, {Damour} T, {Danto} P, {Dittus} H, {Fayet} P, {Foulon} B,
  {Gageant} C, {Guidotti} P~Y, {Hagedorn} D, {Hardy} E, {Huynh} P~A,
  {Inchauspe} H, {Kayser} P, {Lala} S, {L{\"a}mmerzahl} C, {Lebat} V, {Leseur}
  P, {Liorzou} F, {List} M, {L{\"o}ffler} F, {Panet} I, {Pouilloux} B, {Prieur}
  P, {Rebray} A, {Reynaud} S, {Rievers} B, {Robert} A, {Selig} H, {Serron} L,
  {Sumner} T, {Tanguy} N and {Visser} P 2017 {\em Physical Review Letters\/}
  {\bf 119} 231101 (\textit{Preprint} \eprint{1712.01176})

\bibitem{viswanathan18}
{Viswanathan} V, {Fienga} A, {Minazzoli} O, {Bernus} L, {Laskar} J and
  {Gastineau} M 2018 {\em MNRAS\/} {\bf 476} 1877--1888 (\textit{Preprint}
  \eprint{1710.09167})

\bibitem{Uzan:2002vq}
Uzan J~P 2003 {\em Rev. Mod. Phys.\/} {\bf 75} 403 (\textit{Preprint}
  \eprint{hep-ph/0205340})

\bibitem{Uzan:2010pm}
Uzan J~P 2011 {\em Living Rev. Rel.\/} {\bf 14} 2 (\textit{Preprint}
  \eprint{1009.5514})

\bibitem{Uzan:2004qr}
Uzan J~P 2005 {\em AIP Conf. Proc.\/} {\bf 736} 3--20 [,3(2004)]
  (\textit{Preprint} \eprint{astro-ph/0409424})

\bibitem{abbott16}
{Abbott} B~P, {Abbott} R, {Abbott} T~D, {Abernathy} M~R, {Acernese} F, {Ackley}
  K, {Adams} C, {Adams} T, {Addesso} P, {Adhikari} R~X and et~al 2016 {\em
  Physical Review Letters\/} {\bf 116} 061102 (\textit{Preprint}
  \eprint{1602.03837})

\bibitem{abbott17}
{Abbott} B~P, {Abbott} R, {Abbott} T~D, {Acernese} F, {Ackley} K, {Adams} C,
  {Adams} T, {Addesso} P, {Adhikari} R~X, {Adya} V~B and et~al 2017 {\em
  Physical Review Letters\/} {\bf 119} 161101 (\textit{Preprint}
  \eprint{1710.05832})

\bibitem{baker17}
{Baker} T, {Bellini} E, {Ferreira} P~G, {Lagos} M, {Noller} J and {Sawicki} I
  2017 {\em Physical Review Letters\/} {\bf 119} 251301 (\textit{Preprint}
  \eprint{1710.06394})

\bibitem{creminelli17}
{Creminelli} P and {Vernizzi} F 2017 {\em Physical Review Letters\/} {\bf 119}
  251302 (\textit{Preprint} \eprint{1710.05877})

\bibitem{ezquiaga17}
{Ezquiaga} J~M and {Zumalac{\'a}rregui} M 2017 {\em Physical Review Letters\/}
  {\bf 119} 251304 (\textit{Preprint} \eprint{1710.05901})

\bibitem{sakstein17}
{Sakstein} J and {Jain} B 2017 {\em Physical Review Letters\/} {\bf 119} 251303
  (\textit{Preprint} \eprint{1710.05893})

\bibitem{damour92}
{Damour} T and {Esposito-Farese} G 1992 {\em Classical and Quantum Gravity\/}
  {\bf 9} 2093--2176

\bibitem{vainshtein72}
{Vainshtein} A~I 1972 {\em Physics Letters B\/} {\bf 39} 393--394

\bibitem{Damour:1992kf}
Damour T and Nordtvedt K 1993 {\em Phys. Rev. Lett.\/} {\bf 70} 2217--2219

\bibitem{damour94}
{Damour} T and {Polyakov} A~M 1994 {\em Nucl. Phys. B\/} {\bf 423} 532--558
  (\textit{Preprint} \eprint{hep-th/9401069})

\bibitem{khoury04a}
{Khoury} J and {Weltman} A 2004 {\em Phys. Rev. D\/} {\bf 69} 044026
  (\textit{Preprint} \eprint{astro-ph/0309411})

\bibitem{khoury04b}
{Khoury} J and {Weltman} A 2004 {\em Phys. Rev. Lett.\/} {\bf 93} 171104
  (\textit{Preprint} \eprint{astro-ph/0309300})

\bibitem{babichev09}
{Babichev} E, {Deffayet} C and {Ziour} R 2009 {\em Int. J. Mod. Phys. D\/} {\bf
  18} 2147--2154 (\textit{Preprint} \eprint{0905.2943})

\bibitem{hinterbichler10}
{Hinterbichler} K and {Khoury} J 2010 {\em Phys. Rev. Lett.\/} {\bf 104} 231301
  (\textit{Preprint} \eprint{1001.4525})

\bibitem{brax13}
{Brax} P, {Burrage} C and {Davis} A~C 2013 {\em JCAP\/} {\bf 1} 020
  (\textit{Preprint} \eprint{1209.1293})

\bibitem{burrage18}
{Burrage} C and {Sakstein} J 2018 {\em Living Reviews in Relativity\/} {\bf 21}
  1 (\textit{Preprint} \eprint{1709.09071})

\bibitem{berge18}
{Berg{\'e}} J, {Brax} P, {M{\'e}tris} G, {Pernot-Borr{\`a}s} M, {Touboul} P and
  {Uzan} J~P 2018 {\em Physical Review Letters\/} {\bf 120} 141101
  (\textit{Preprint} \eprint{1712.00483})

\bibitem{fischbach99}
{Fischbach} E and {Talmadge} C~L 1999 {\em {The Search for Non-Newtonian
  Gravity}\/}

\bibitem{kapner07}
{Kapner} D~J, {Cook} T~S, {Adelberger} E~G, {Gundlach} J~H, {Heckel} B~R,
  {Hoyle} C~D and {Swanson} H~E 2007 {\em Physical Review Letters\/} {\bf 98}
  021101 (\textit{Preprint} \eprint{hep-ph/0611184})

\bibitem{masuda09}
{Masuda} M and {Sasaki} M 2009 {\em Physical Review Letters\/} {\bf 102} 171101
  (\textit{Preprint} \eprint{0904.1834})

\bibitem{sushkov11}
{Sushkov} A~O, {Kim} W~J, {Dalvit} D~A~R and {Lamoreaux} S~K 2011 {\em Physical
  Review Letters\/} {\bf 107} 171101 (\textit{Preprint} \eprint{1108.2547})

\bibitem{klimchitskaya14}
{Klimchitskaya} G~L and {Mostepanenko} V~M 2014 {\em Gravitation and
  Cosmology\/} {\bf 20} 3--9 (\textit{Preprint} \eprint{1403.5778})

\bibitem{rummel11}
{Rummel} R, {Yi} W and {Stummer} C 2011 {\em Journal of Geodesy\/} {\bf 85}
  777--790

\bibitem{pail11}
{Pail} R, {Bruinsma} S, {Migliaccio} F, {F{\"o}rste} C, {Goiginger} H, {Schuh}
  W~D, {H{\"o}ck} E, {Reguzzoni} M, {Brockmann} J~M, {Abrikosov} O, {Veicherts}
  M, {Fecher} T, {Mayrhofer} R, {Krasbutter} I, {Sans{\`o}} F and {Tscherning}
  C~C 2011 {\em Journal of Geodesy\/} {\bf 85} 819--843

\bibitem{tapley04}
{Tapley} B~D, {Bettadpur} S, {Watkins} M and {Reigber} C 2004 {\em Geophysical
  Research Letters\/} {\bf 31} L09607

\bibitem{tapley05}
{Tapley} B, {Ries} J, {Bettadpur} S, {Chambers} D, {Cheng} M, {Condi} F,
  {Gunter} B, {Kang} Z, {Nagel} P, {Pastor} R, {Pekker} T, {Poole} S and {Wang}
  F 2005 {\em Journal of Geodesy\/} {\bf 79} 467--478

\bibitem{reigber05}
{Reigber} C, {Schmidt} R, {Flechtner} F, {K{\"o}nig} R, {Meyer} U, {Neumayer}
  K~H, {Schwintzer} P and {Zhu} S~Y 2005 {\em Journal of Geodynamics\/} {\bf
  39} 1--10

\bibitem{pail10}
{Pail} R, {Goiginger} H, {Schuh} W~D, {H{\"o}ck} E, {Brockmann} J~M, {Fecher}
  T, {Gruber} T, {Mayer-G{\"u}rr} T, {Kusche} J, {J{\"a}ggi} A and {Rieser} D
  2010 {\em Geophysical Research Letters\/} {\bf 37} L20314

\bibitem{mayer06}
Mayer-Gurr T, Eicker A and Ilk K~H 2006 {\em Proc. First Symp. Int. Grav. Field
  Ser.\/}

\bibitem{pavlis12}
{Pavlis} N~K, {Holmes} S~A, {Kenyon} S~C and {Factor} J~K 2012 {\em Journal of
  Geophysical Research (Solid Earth)\/} {\bf 117} B04406

\bibitem{hoyle04}
{Hoyle} C~D, {Kapner} D~J, {Heckel} B~R, {Adelberger} E~G, {Gundlach} J~H,
  {Schmidt} U and {Swanson} H~E 2004 {\em Phys. Rev. D\/} {\bf 70} 042004
  (\textit{Preprint} \eprint{hep-ph/0405262})

\bibitem{fischbach86}
{Fischbach} E, {Sudarsky} D, {Szafer} A, {Talmadge} C and {Aronson} S~H 1986
  {\em Physical Review Letters\/} {\bf 56} 3--6

\bibitem{toth18}
{T{\'o}th} G 2018 {\em ArXiv e-prints\/} (\textit{Preprint}
  \eprint{1803.04720})

\bibitem{ciufolini96}
{Ciufolini} I 1996 {\em Nuovo Cimento A Serie\/} {\bf 109} 1709--1720

\bibitem{metzler05}
{Metzler} B and {Pail} R 2005 {\em Studia Geophysica et Geodaetica\/} {\bf 49}
  441--462

\bibitem{wagnermcadoo12}
{Wagner} C~A and {McAdoo} D~C 2012 {\em Journal of Geodesy\/} {\bf 86} 99--108

\bibitem{shako14}
Shako R, Förste C, Abrykosov O, Bruinsma S, Marty J~C, Lemoine J~M, Flechtner
  F, Neumayer K~H and Dahle C 2014 {\em EIGEN-6C: A High-Resolution Global
  Gravity Combination Model Including GOCE Data\/} 1st ed (GEOTECHNOLOGIEN
  Science Report; No. 20; Advanced Technologies in Earth Sciences, Berlin
  [u.a.]: Springer, 155-161) ISBN 978-3-642-32134-4, 978-3-642-32135-1

\bibitem{abramowitz+stegun}
Abramowitz M and Stegun I~A 1964 {\em Handbook of Mathematical Functions with
  Formulas, Graphs, and Mathematical Tables\/} ninth dover printing, tenth gpo
  printing ed (New York: Dover)

\bibitem{cunningham70}
{Cunningham} L~E 1970 {\em Celestial Mechanics\/} {\bf 2} 207--216

\bibitem{metris98}
{M{\'e}tris} G, {Xu} J and {Wytrzyszczak} I 1998 {\em Celestial Mechanics and
  Dynamical Astronomy\/} {\bf 71} 137--151

\bibitem{fantino09}
{Fantino} E and {Casotto} S 2009 {\em Journal of Geodesy\/} {\bf 83} 595--619

\bibitem{petrovskaya10}
{Petrovskaya} M~S and {Vershkov} A~N 2010 {\em Journal of Geodesy\/} {\bf 84}
  165--178

\bibitem{newman66}
{Newman} E~T and {Penrose} R 1966 {\em Journal of Mathematical Physics\/} {\bf
  7} 863--870

\bibitem{goldberg67}
{Goldberg} J~N, {Macfarlane} A~J, {Newman} E~T, {Rohrlich} F and {Sudarshan}
  E~C~G 1967 {\em Journal of Mathematical Physics\/} {\bf 8} 2155--2161

\bibitem{kaula66}
Kaula W~M 1966 {\em Theory of Satellite Geodesy: Applications of Satellites to
  Geodesy\/} 1st ed (Dover Publications Inc.) ISBN 0486414655, 978-0486414652

\bibitem{uzan}
Deruelle N and Uzan J~P 2014 {\em Th\'eories de la Relativit\'e\/} 1st ed
  (Belin) ISBN 2701158486, 978-2701158488

\bibitem{roy05}
{Roy} A~E 2005 {\em {Orbital motion}\/} 4th ed (Bristol (UK): Institute of
  Physics Publishing) ISBN 0-7503-1015-6

\bibitem{montenbruck}
Montenbruck O and Gill E 2000 {\em Satellite Orbits: Models, Methods, and
  Applications\/} Physics and astronomy online library (Springer Berlin
  Heidelberg) ISBN 9783540672807

\bibitem{iorio02b}
{Iorio} L 2002 {\em Physics Letters A\/} {\bf 298} 315--318 (\textit{Preprint}
  \eprint{gr-qc/0201081})

\bibitem{haranas11b}
{Haranas} I and {Ragos} O 2011 {\em Astrophys. Space Sci\/} {\bf 331} 115--119

\bibitem{haranas11a}
{Haranas} I, {Ragos} O and {Mioc} V 2011 {\em Astrophys. Space Sci\/} {\bf 332}
  107--113

\bibitem{kolosnitsyn04}
{Kolosnitsyn} N~I and {Melnikov} V~N 2004 {\em General Relativity and
  Gravitation\/} {\bf 36} 1619--1624 (\textit{Preprint} \eprint{gr-qc/0302048})

\bibitem{haranas16}
{Haranas} I, {Kotsireas} I, {G{\'o}mez} G, {Fullana} M~J and {Gkigkitzis} I
  2016 {\em Astrophys. Space Sci\/} {\bf 361} 365

\bibitem{lucchesi11}
{Lucchesi} D~M 2011 {\em Advances in Space Research\/} {\bf 47} 1232--1237

\bibitem{hagiwara89}
{Hagiwara} Y 1989 {\em Journal of the Geodesic Society of Japan\/} {\bf 35}
  319--324

\bibitem{grombein13}
{Grombein} T, {Seitz} K and {Heck} B 2013 {\em Journal of Geodesy\/} {\bf 87}
  645--660

\bibitem{casenave16}
{Casenave} F, {M{\'e}tivier} L, {Pajot-M{\'e}tivier} G and {Panet} I 2016 {\em
  Journal of Geodesy\/} {\bf 90} 655--675

\bibitem{trotta08}
{Trotta} R 2008 {\em Contemporary Physics\/} {\bf 49} 71--104
  (\textit{Preprint} \eprint{0803.4089})

\bibitem{liddle04}
{Liddle} A~R 2004 {\em Monthly Notices of the Royal Astronomical Society\/}
  {\bf 351} L49--L53 (\textit{Preprint} \eprint{astro-ph/0401198})

\bibitem{seehars16}
{Seehars} S, {Grandis} S, {Amara} A and {Refregier} A 2016 {\em Phys. Rev. D\/}
  {\bf 93} 103507 (\textit{Preprint} \eprint{1510.08483})

\bibitem{raveri18}
{Raveri} M and {Hu} W 2018 {\em ArXiv e-prints\/} (\textit{Preprint}
  \eprint{1806.04649})

\bibitem{cuyt}
Cuyt A~A, Petersen V, Verdonk B, Waadeland H and Jones W~B 2008 {\em Handbook
  of Continued Fractions for Special Functions\/} 1st ed (Springer Publishing
  Company, Incorporated) ISBN 1402069480, 9781402069482

\end{thebibliography}

\end{document}